\documentclass[final,3p,11pt,times]{elsarticle}
\biboptions{sort&compress}

\usepackage{amssymb}
\usepackage{amsmath}

\usepackage{upgreek}
\usepackage{natbib}
\usepackage{comment}
\usepackage{enumitem}

\usepackage{bm}
\newcommand{\vct}[1]{\bm{\mathrm{#1}}} 
\newcommand{\ten}[1]{\bm{\mathrm{#1}}} 

\usepackage[colorlinks=true,
            linkcolor=blue,
            citecolor=blue,
            urlcolor=blue]{hyperref}

\usepackage[nameinlink,noabbrev]{cleveref}

\crefname{figure}{Fig.}{Figs.}
\crefname{equation}{}{}
\crefname{section}{Section}{Sections}
\crefname{table}{Table}{Tables}
\crefname{algorithm}{Algorithm}{Algorithms}

\usepackage{graphicx}
\usepackage{algorithm}
\usepackage{algpseudocode}
\usepackage{tikz}
\usetikzlibrary{arrows.meta,positioning,shapes.geometric,shapes.misc,fit}
\pgfdeclarelayer{background}
\pgfsetlayers{background,main}
\usepackage[dvipsnames]{xcolor}
\usepackage{soul}
\sethlcolor{yellow}
\soulregister\cref7
\soulregister\ref7
\soulregister\citep7
\soulregister\citet7

\begin{document}

\begin{frontmatter}



\title{Data-efficient Bayesian-guided design selection from large candidate sets: Application to hyperelastic stochastic metamaterials} 


\author[]{Hooman Danesh\corref{cor1}} 
\ead{hooman.danesh@tu-braunschweig.de}
\cortext[cor1]{Corresponding author}

\author{Henning Wessels}

\affiliation{organization={Division of Data-Driven Modeling of Mechanical Systems, Institute of Applied Mechanics, Technische Universität Braunschweig},
            addressline={Pockelsstr. 3}, 
            city={38106 Braunschweig},
            country={Germany}}

\begin{abstract}

From a pool of admissible designs, we aim to identify a structure that achieves a target macroscopic stress response. For each candidate, the response is obtained from a high-fidelity oracle, such as expensive computational homogenization or experiments. We consider cases in which (i) the geometry cannot be conveniently parameterized, rendering gradient-based optimization inapplicable, and (ii) brute-force evaluation of all candidates is infeasible due to costly oracle queries. To tackle this challenge, we propose a Bayesian-guided design selection framework. The dimensionality of design variants is reduced through statistical feature engineering, and the resulting low-dimensional descriptors are mapped to effective hyperelastic constitutive parameters using a multi-output Gaussian process surrogate. The surrogate is trained using uncertainty-driven active learning with only a limited number of high-fidelity oracle evaluations. The surrogate shortlists promising candidates, and since its accuracy is inherently limited, the final selection of the optimal design is performed through high-fidelity oracle evaluations within the shortlist. In numerical test cases, we consider a design set of 50,000 candidate structures. Active learning requires labeling less than half a percent of the entire candidate set. Bayesian-guided design selection reaches a prescribed error threshold with only a handful of oracle evaluations in most cases.

\end{abstract}




\end{frontmatter}



\section{Introduction}
\label{sec:intro}

Understanding and predicting the behavior of complex systems has long been a central objective across scientific and engineering disciplines. Considerable effort has therefore been devoted to developing accurate physical models and advanced experimental techniques that enable the efficient solution of \emph{forward problems}, in which the response of a system is computed for a given set of input parameters or input functions. In contrast, the corresponding \emph{inverse problems}, which aim to determine the input parameters required to achieve a desired target response, remain challenging to solve. The main difficulty arises from the fact that inverse problems typically require a large number of forward evaluations in order to identify suitable solutions. Such evaluations, whether obtained from high-fidelity numerical simulations or experimental measurements, are often computationally or experimentally expensive. Consequently, it is highly desirable to develop inverse frameworks that can identify optimal designs while requiring as few forward evaluations as possible.

Traditional approaches to inverse design often rely on gradient-based optimization, where an objective function quantifying the mismatch between the obtained and desired response is minimized. While effective in many applications, these methods require the gradients of the system outputs with respect to the design parameters to be readily computable. In practice, this requirement is not always satisfied. Many engineering design problems involve complex geometries or discrete design variables that cannot be easily parameterized, making gradient-based methods difficult to apply. Furthermore, the enormous size of the design space often makes exhaustive exploration through brute-force evaluation infeasible. To address these challenges, several computational approaches have been proposed in recent years.

\paragraph{\textbf{Surrogate models for inverse design}} A common strategy for addressing expensive forward evaluations is the construction of surrogate models that approximate the underlying physics. Once trained, such models can rapidly predict system responses, enabling efficient exploration of the design space without repeatedly invoking high-fidelity simulations or experiments. Surrogate-based inverse design has been successfully applied across a wide range of architected material systems. For example, machine learning surrogates such as random forests have been used to approximate the mechanical response of parametrized auxetic unit cells \citep{danesh2024fft}, while physics-augmented neural networks have been coupled with evolutionary strategies to recover anisotropic microstructural parameters \citep{jadoon2025inverse}. Deep learning approaches have also enabled mappings from desired mechanical responses to structural architectures, including physics-constrained neural networks for truss metamaterials \citep{bastek2022truss}, tandem neural networks for spinodoid microstructures \citep{kumar2020inverse}, and artificial neural networks linking target stress–strain curves to semi-auxetic microstructures \citep{mohammadnejad2024ann}. More recently, surrogate models have been extended to geometry-rich representations, where the design variables correspond directly to structural images or graphs. Convolutional neural networks (CNNs) have been used for direct inverse prediction of layered thin film materials \citep{lininger2021thinfilmcnn}, while graph-based learning approaches such as graph neural networks (GNNs) have been employed to capture connectivity in helicoidal laminates \citep{garg2026gcnhelicoidal}, irregular lattice materials \citep{dold2023differentiable}, and nonlinear truss metamaterials \citep{maurizi2025graphmetamat}.

Despite its advantages, surrogate-based inverse design does not by itself eliminate the central bottleneck of the problem: the surrogate must first be trained on sufficiently informative data. In practice, this often requires a large number of high-fidelity evaluations to cover the relevant design space with adequate accuracy, particularly when the input consists of complex microstructures rather than a few low-dimensional parameters. This trade-off is evident across recent applications, where the speed gained at inference time is preceded by a nontrivial cost in generating training data. As a result, surrogate models are highly effective once available, but their construction can remain expensive when the accessible design space is large, high-dimensional, or only weakly parameterized.

\paragraph{\textbf{Bayesian optimization for data-efficient design}} Bayesian optimization addresses the data-efficiency issue more directly by coupling a probabilistic surrogate, often a Gaussian process (GP), with an acquisition function that balances exploration and exploitation. This framework has been successfully applied to a variety of inverse design problems in materials science. Examples include the design of spinodoid metamaterials within parameterized spaces \citep{rassloff2025inverse}, optimization of Bézier-parameterized auxetic unit cells with respect to mechanical performance and fatigue life \citep{kang2025auxetic}, inference of nonlinear interface laws in bio-inspired composites from target stress–strain curves \citep{zhang2025bio}, complex block copolymer systems \citep{dong2023blockcopolymerbo}, and discovery of process–microstructure relationships through batch Bayesian optimization \citep{honarmandi2022batchbo}.

The primary strength of Bayesian optimization lies in its ability to locate optimal solutions using relatively few high-fidelity evaluations. However, its effectiveness is closely linked to its sequential and target-specific nature. In a typical workflow, the surrogate model is constructed and iteratively refined for a specific objective or target response. The surrogate therefore evolves specifically to guide the search toward this desired target response. When the target response changes, the optimization procedure must typically be restarted, requiring the construction of a new surrogate model and the repetition of the associated hyperparameter optimization. This process can become computationally demanding, particularly when the surrogate must be repeatedly rebuilt and refined for multiple inverse design queries. As a result, Bayesian optimization is particularly effective when addressing individual inverse design tasks under limited evaluation budgets, but it becomes less attractive in scenarios where many target responses must be explored within the same design space.

\paragraph{\textbf{Generative models for design exploration}} When the design variables correspond to complex geometries rather than low-dimensional parameter sets, inverse design becomes considerably more challenging. To address this issue, several studies have proposed generative modeling approaches that learn a low-dimensional latent representation of the design space. Optimization or sampling can then be performed in this latent space to generate candidate structures with desired properties. Examples include generative models for microstructure generation and inverse design based on variational autoencoders \citep{wang2020deep,ma2019probabilistic,attari2023pfvae,zheng2023truss,nguyen2026deep,zhang2025generative}, as well as normalizing flows for design of porous and stochastic microstructures \citep{mirzaee2025inverse,zang2026designgenno}. Variational autoencoders combined with Bayesian optimization have further been used to explore latent spaces for designing manufacturable phononic metaplates \citep{fan2026metaplate}. Bayesian generative formulations combining latent representations with GP likelihood models have likewise been explored \citep{generale2024inverse}. Generative models have also been integrated with active learning strategies for designing polycrystalline textures with target anisotropic responses \citep{buzzy2025active}. A related but distinct line of work employs diffusion-based generative models rather than explicit latent space optimization; for instance, conditional diffusion has been used to generate level-set-parameterized lattice unit cells from target elastic properties \citep{zhang2025latticeoptdiff}, while video denoising diffusion models have been used to generate metamaterial microstructures from nonlinear target stress--strain responses \citep{bastek2023inverse}.

These approaches provide a powerful mechanism for exploring high-dimensional geometric design spaces without requiring explicit parameterization of structures. However, they also introduce several practical challenges. Training generative models typically requires large and representative labeled datasets in order to learn meaningful latent representations. In addition, the structures generated in latent space are not always guaranteed to satisfy manufacturability constraints or geometric integrity unless these constraints are explicitly embedded into the learning process. In many cases, additional reconstruction or post-processing steps are therefore required to obtain physically realizable geometries.

\paragraph{\textbf{Our Bayesian-guided design selection framework}} Motivated by the limitations of existing inverse design strategies, we propose a data-efficient framework for candidate screening and design selection under a limited budget of high-fidelity evaluations. A schematic overview of the proposed workflow is shown in \cref{fig:concept}. In essence, a data-efficient surrogate model is used to screen a large pool of candidate designs and identify a shortlist of promising structures, after which high-fidelity physics-based evaluations determine the final design. Rather than performing optimization in a continuous latent space, the proposed approach operates on a predefined pool of admissible candidate structures. This perspective is particularly suitable for problems where microstructural geometries cannot be conveniently parameterized and exhaustive evaluation of all candidates using high-fidelity evaluations is impractical. The framework is built upon the following key components:

\begin{figure}[h]
\centering
\includegraphics{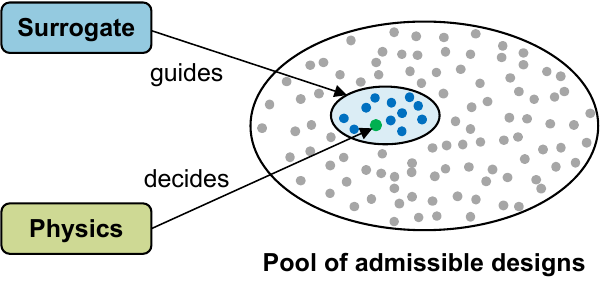}
\caption{Schematic overview of the proposed Bayesian-guided design selection framework. A data-efficient surrogate model guides the exploration of a large candidate design set toward a shortlist of promising structures, while high-fidelity physics-based evaluations ultimately decide the final design.}
\label{fig:concept}
\end{figure}

\begin{itemize}
    \item \textbf{Discrete design space formulation.}  
    The inverse problem is formulated as a \emph{selection task} over a finite library of candidate structures. Since the design pool is defined \emph{a priori}, geometric feasibility and manufacturability are guaranteed by construction. This avoids the need for microstructure reconstruction or generative modeling, which can introduce additional complexity and may not always guarantee physically realizable designs.
    \item \textbf{Compact structural feature representation.}  
    As raw microstructure images are inherently high-dimensional, statistical feature engineering is employed to extract compact descriptors that retain the essential structural information. This representation enables efficient surrogate learning while avoiding the need to train complex neural networks to learn geometric embeddings or latent representations of the design space.
    \item \textbf{Probabilistic surrogate modeling with uncertainty-driven active learning.}  A probabilistic surrogate model is constructed to approximate the structure--property relationship using only a small number of labeled samples. To minimize the required high-fidelity evaluations, uncertainty-driven active learning is employed, in which the predictive uncertainty of the surrogate guides the selection of new candidate structures whose evaluations provide the greatest information gain. In contrast to Bayesian optimization frameworks, where a surrogate model is typically constructed and iteratively refined for a specific target during the optimization process, the surrogate in the present framework is trained only once to represent the entire admissible design space. Once constructed, it can therefore be reused to support multiple inverse queries without requiring repeated surrogate training or hyperparameter optimization.
    \item \textbf{Uncertainty-aware candidate screening.}  
    Once trained, the surrogate efficiently evaluates the candidate pool during the design selection stage. Promising structures are identified based on their predicted proximity to the target response, while predictive uncertainty is incorporated into the ranking to penalize unreliable predictions. In contrast to gradient-based optimization methods, which may converge to local minima in complex design landscapes, this screening approach enables robust exploration of the discrete design space.
    \item \textbf{Physics-based final decision.}  
    The final design is always determined through a small number of high-fidelity oracle evaluations. In this way, the surrogate \emph{guides the search}, while the governing physics ultimately \emph{decides the final design}. This separation between surrogate guidance and physics-based validation leads to a non-intrusive framework in which the machine learning model does not need to explicitly enforce physical constraints or governing equations.
\end{itemize}

 The proposed framework is applied to hyperelastic microstructures, where the objective is to identify a structure whose effective macroscopic stress response matches a prescribed target behavior. In this setting, the design space consists of a large pool of admissible stochastic metamaterials, from which the framework efficiently identifies candidates that achieve the desired response using only a small number of high-fidelity oracle evaluations.

\section{Bayesian-guided design selection}
\label{sec:inv_des}

In many applications, the admissible design space is known \emph{a priori}, for instance, through combinatorial construction rules, manufacturing constraints, or a precomputed database of representative designs. This motivates candidate selection formulation in which the goal is to \emph{select} the best structure from a finite set of candidate structures $\mathcal{D}=\{\vct{M}_i\}_{i=1}^{n_d}$ with the index set $\mathcal{I}=\{1,\dots,n_d\}$, where $n_d$ is the total number of designs. Throughout this contribution, but without loss of generality, each design $\vct{M}_i\in\mathbb{R}^{n_p}$ is a discretized microstructure represented as a 2D/3D bitmap with $n_p$ pixels/voxels. The design selection task is to identify the structure $ \vct{M}^\star$, 
\begin{equation}
    \vct{M}^\star
    =
    \mathop{\arg\min}\limits_{\vct{M}_i \in \mathcal{D}}
    \mathcal{L}_i,
    \label{eq:opt_1}
\end{equation}
whose response best matches a prescribed target first Piola--Kirchhoff stress response. To define this objective, let $\ten{P}^{\star}\in\mathbb{R}^{3\times3}$ denote the target stress response over a prescribed set of deformation gradients $\mathcal{F}=\{\ten{F}_j\}_{j=1}^{n_f}$ with $\ten{F}_j\in\mathbb{R}^{3\times3}$, where $n_f$ is the total number of deformation states. The loss $\mathcal{L}_i$ is then defined as
\begin{equation}
    \mathcal{L}_{i}
    =
    \frac{1}{n_f}
    \sum_{j=1}^{n_f}
    \left\|
    \mathcal{S}_{\mathrm{tar}}\Big(\ten{P}^{\star}\big(\ten{F}_j\big)\Big)
    -
    \mathcal{S}_{\mathrm{tar}}\Big(\ten{P}\big(\ten{F}_j;\vct{M}_i\big)\Big)
    \right\|_2^2,
    \qquad
    \forall\, i\in\mathcal{I},
    \label{eq:inv_loss}
\end{equation}
which averages the squared stress mismatch across the prescribed deformation set and provides a single scalar score for each discrete candidate. Here, $\ten{P}(\ten{F}_j;\vct{M}_i)\in\mathbb{R}^{3\times3}$ is the effective homogenized first Piola--Kirchhoff stress response, and $\mathcal S_{\mathrm{tar}}:\mathbb{R}^{3\times3}\to\mathbb{R}^{n_{\mathrm{tar}}}$ extracts, scales, and vectorizes the active stress components chosen as targets, with $n_{\mathrm{tar}}$ the number of such target components. The oracle $\mathcal{O}$ provides the high-fidelity homogenized response, typically through experiments or physics-based simulations:
\begin{equation}
    \ten{P}\left(\ten{F}_j;\vct{M}_i\right)
    =
    \mathcal{O}\left(\ten{F}_j;\vct{M}_i\right),
    \qquad
    \mathcal{O}: \mathcal{D} \times \mathbb{R}^{3\times3}\rightarrow \mathbb{R}^{3\times3}.
    \label{eq:oracle}
\end{equation}
An ad-hoc procedure for the identification of $\vct{M}^\star$ from \cref{eq:opt_1}  is to evaluate and rank the loss $\mathcal{L}_i$ for each design $\vct{M}_i$. Over the full design set $\mathcal{D}$, this entails a prohibitively large number of oracle evaluations and is generally infeasible.

\paragraph{\textbf{Effective macroscopic constitutive model}} We therefore assume that the effective homogenized stress $\ten{P}$ can be approximated by an effective constitutive response $\widetilde{\ten{P}}\in\mathbb{R}^{3\times3}$. We postulate a stochastic observation model,
\begin{equation}
    \ten{P}\left(\ten{F}_j;\vct{M}_i\right)
    =
    \widetilde{\ten{P}}\left(\ten{F}_j;\vct{\uptheta}_i\right) + \vct{\varepsilon},
    \label{eq:effective_stress}
\end{equation}
Here, $\vct{\varepsilon}\in\mathbb{R}^{3\times3}$ is an error term representing the discrepancy between the oracle response and its effective constitutive approximation. For a numerical oracle, this term accounts for numerical errors and model-form discrepancy, whereas for an experimental oracle it additionally captures measurement uncertainty. Its probabilistic specification is introduced in \cref{sec:obs_model}. Moreover,  $\vct{\uptheta}_i\in\mathbb{R}^{n_\theta}$ denotes a vector of random variables, namely the effective material parameters for the $i$-th design $\vct{M}_i$. 

Following first principles, we postulate the existence of a scalar-valued  hyperelastic strain energy density $\widetilde{W}(\ten{F}_j;\vct{\uptheta}_i)$, whose derivative yields the stress response. Following the notation of \citep{flaschel2021euclid}, the effective strain energy density can be written as a linear combination:
\begin{equation}
    \widetilde{W}\left(\ten{F}_j;\vct{\uptheta}_i\right)
    =
    \vct{Q}\left(\ten{F}_j\right)\cdot \vct{\uptheta}_i
    -
    p\,\left(J-1\right).
    \label{eq:energy_linear}
\end{equation}
Here, $\vct{Q}(\ten{F})\in\mathbb{R}^{n_\theta}$ collects the chosen energy basis functions (e.g., functions of invariants or principal stretches), $J=\det\ten{F}$ denotes the Jacobian, and $p$ is the hydrostatic pressure acting as a Lagrange multiplier enforcing incompressibility. The effective stress tensor is then obtained by differentiation:
\begin{equation}
    \widetilde{\ten{P}}\left(\ten{F}_j;\vct{\uptheta}_i\right)
    =
    \left.
    \frac{\partial \widetilde{W}\left(\ten{F};\vct{\uptheta}_i\right)}{\partial \ten{F}}
    \right|_{\ten{F}=\ten{F}_j}
    =
    \left.
    \frac{\partial \vct{Q}\left(\ten{F}\right)}{\partial \ten{F}}
    \right|_{\ten{F}=\ten{F}_j}
    \cdot
    \vct{\uptheta}_i
    -
    p\,\ten{F}_j^{-\top}.
    \label{eq:eff_stress}
\end{equation}
Identification of the constitutive model structure and parameters $\vct{\uptheta}_i$ satisfying  \cref{eq:effective_stress} constitutes a constitutive model calibration or discovery task, which again relies on high-fidelity oracle evaluations. To reduce the computational cost, we are ultimately interested in a surrogate for the map $\vct{M}_i \mapsto \vct{\uptheta}_i$. To enforce the positivity of material parameters, we consider the componentwise transformation
\begin{equation}
    \vct{\uptheta}_i
    =
    \log \Big(1 + \exp\big(\vct{\upxi}_i\big)\Big).
    \label{eq:pos_map}
\end{equation}
Here, we introduce $\vct{\xi}_i\in\mathbb{R}^{n_\theta}$ as a latent random variable, for which we train a GP surrogate using variational inference and active learning, minimizing the required oracle evaluations. Details can be found in \cref{sec:gp_surrogate}.

\paragraph{\textbf{Shortlisting}} Building on the trained probabilistic surrogate, we approximate the expensive oracle mismatch \cref{eq:inv_loss} by a surrogate mismatch,
\begin{equation}
    \widetilde{\mathcal L}_i
    =
    \frac{1}{n_f}
    \sum_{j=1}^{n_f}
    \left\|
    \mathcal{S}_{\mathrm{tar}}\Big(\ten{P}^{\star}\big(\ten{F}_j\big)\Big)
    -
    \mathcal{S}_{\mathrm{tar}}\Big(\widetilde{\ten{P}}\big(\ten{F}_j;\vct{\uptheta}_i(\vct{\xi}_i)\big)\Big)
    \right\|_2^2,
    \qquad
    \forall\, i\in\mathcal{I},
    \label{eq:inv_loss_surrogate}
\end{equation}
which allows us to efficiently screen the full design set. Since $\vct{\xi}_i$ is a random variable, the pushforward measure $\widetilde{\mathcal L}_i$ is random as well. Therefore, we first screen the full design set using a point estimate $\widehat{\vct{\uptheta}}_i$ obtained from passing the surrogate's predictive mean $\mathbb{E}[\vct{\upxi}_i]$ through the componentwise positivity map \cref{eq:pos_map}:
\begin{equation}
    \widehat{\vct{\uptheta}}_i
    =
    \log\Big(
    1 + \exp \big(\mathbb{E}\left[\vct{\upxi}_i\right]\big)
    \Big).
    \label{eq:predictive_mean}
\end{equation}
Note that, due to the nonlinearity of the positivity map \cref{eq:pos_map}, in general $\mathbb{E}[\vct{\uptheta}_i] \neq \log(1 + \exp(\mathbb{E}[\vct{\upxi}_i]))$, and the expectation $\mathbb{E}[\vct{\uptheta}_i]$ and its pushforward measures, such as the surrogate mismatch \cref{eq:inv_loss_surrogate}, are typically estimated via Monte Carlo sampling. Nevertheless, the point estimate \cref{eq:predictive_mean} provides a computationally inexpensive proxy for screening the full design set. Using this point estimate, we then evaluate the loss function for all candidates in the design set and retain a shortlist of size $E_{\max}$,
\begin{equation}
    \mathcal{B}^\star
    =
    \mathop{\arg\min}\limits_{\substack{\mathcal B \subset \mathcal{I}\\|\mathcal B|=E_{\max}}}
    \left.
    \widetilde{\mathcal L}_i
    \right|_{\vct{\uptheta}_i = \widehat{\vct{\uptheta}}_i}, 
    \qquad
    \forall\, i\in\mathcal{I}, 
    \label{eq:inv_shortlist}
\end{equation}
with $\mathcal{B}^\star\subset\mathcal{I}$ the set of indices for the most promising candidates according to the surrogate mean.

\paragraph{\textbf{Uncertainty-aware scoring}} The decision on whether a design is shortlisted depends solely on the surrogate's predictive mean through the point estimate $\widehat{\vct{\uptheta}}_i$ defined in \cref{eq:predictive_mean}. 
We therefore compute, for all shortlisted designs $\mathcal{B}^\star$, an uncertainty-aware score $\phi_i$ that takes into account Monte Carlo estimates of both the mean and variance of the surrogate mismatch \cref{eq:inv_loss_surrogate}:
\begin{equation}
    \phi_i =
    \mathbb{E}\left[\widetilde{\mathcal L}_i\right]
    +
    \lambda\,
    \sqrt{\mathrm{Var}\left[\widetilde{\mathcal L}_i\right]},
    \qquad
    \forall\, i\in\mathcal{B^{\star}}.
    \label{eq:score}
\end{equation}
The variance, weighted by $\lambda\ge0$, penalizes candidates for which the surrogate prediction is accompanied by high epistemic uncertainty (cf. \cref{sec:gp_surrogate}). The risk parameter $\lambda$ controls the influence of predictive uncertainty in the selection rule: when $\lambda=0$, the rule reduces to the pure exploitation of the surrogate mean, and when $\lambda \to \infty$, the selection becomes dominated by variance minimization, favoring the most confident predictions. To obtain a scale-consistent and problem-adaptive choice, we set $\lambda$ based on the statistics of the shortlist $\mathcal{B}^\star$:
\begin{equation}
    \lambda
    =
    \frac{\overline{\mu}}{\overline{\sigma}},
    \qquad
    \overline{\mu}=\frac{1}{|\mathcal{B}^\star|}\sum_{i\in\mathcal{B}^\star}
    \mathbb{E}\left[\widetilde{\mathcal L}_i\right],
    \qquad
    \overline{\sigma}=\frac{1}{|\mathcal{B}^\star|}\sum_{i\in\mathcal{B}^\star}
    \sqrt{\mathrm{Var}\left[\widetilde{\mathcal L}_i\right]}.
\end{equation}
This normalization balances the typical magnitude of the mean and variance terms within the shortlist, ensuring that neither component dominates the score purely due to scaling.

\paragraph{\textbf{Oracle evaluations}} Recall that the focus of the presented methodology is to minimize the number of oracle evaluations needed for design selection. This premise motivates the postulation of an effective macroscopic constitutive model \cref{eq:effective_stress} and the surrogate modeling strategy outlined in \cref{sec:gp_surrogate}. Consequently, our ranking based on the surrogate mismatch \cref{eq:inv_loss_surrogate} is an approximation that must only be used to screen the design set $\mathcal{D}$ for potential candidates retained in the shortlist $\mathcal{B}^\star$. The final selection of a design, i.e., the approximate solution of \cref{eq:opt_1}, must be based on the oracle mismatch \cref{eq:inv_loss}. Thus, we successively evaluate the oracle \cref{eq:oracle} for the shortlisted designs $\mathcal{B}^\star$, starting with the design associated with the minimum score \cref{eq:score}, 
\begin{equation}
    i_e^\star
    =
    \mathop{\arg\min}\limits_{i \in \mathcal{B}^\star\setminus\mathcal{Q}_{e-1}} \phi_i,
    \qquad
    \text{with }
    \mathcal{Q}_e = \mathcal{Q}_{e-1}\cup\{i_e^\star\}.
    \label{eq:inv_select}
\end{equation}
Here, $e\in\{1,\dots,E_{\max}\}$ denotes the iteration index, and $\mathcal{Q}_e$ is the set of design indices evaluated by the oracle up to iteration $e$.

\paragraph{\textbf{Termination threshold}}  As evaluating the oracle across the entire shortlist would be unnecessarily costly, we use a stopping rule that allows early termination once adequate agreement with the target response is achieved. To quantify this agreement, we define a normalized mean absolute error (nMAE) for each target stress component across all sampled deformation states:
\begin{equation}
    \mathrm{nMAE}_{p}\left(\vct{M}_{i_e^\star}\right)
    =
    \frac{
    \frac{1}{n_f}\sum_{j=1}^{n_f}
    \left|
    \mathcal{S}_{\mathrm{tar}}\Big(\ten{P}^{\star}\big(\ten{F}_j\big)\Big)_{p}
    -
    \mathcal{S}_{\mathrm{tar}}\Big(\ten{P}\big(\ten{F}_j;\vct{M}_{i_e^\star}\big)\Big)_{p}
    \right|
    }{
    \frac{1}{n_f}\sum_{j=1}^{n_f}
    \left|
    \mathcal{S}_{\mathrm{tar}}\Big(\ten{P}^{\star}\big(\ten{F}_j\big)\Big)_{p}
    \right|
    },
    \qquad
    p = 1,\dots, n_\mathrm{tar}.
    \label{eq:inv_nmae_comp}
\end{equation}
The componentwise nMAE is a discrete normalized $L_1$ distance between the target and oracle stress responses over the sampled deformation states. The $L_1$ norm measures the average relative mismatch over the sampled deformation states, giving $\eta$ a direct engineering interpretation while limiting the influence of isolated large deviations. When greater sensitivity to large deviations or worst-case control is required, normalized $L_2$ or $L_\infty$ measures, respectively, can be used instead without changing the design selection procedure. The normalization by the target magnitude yields comparable errors across selected target components and different loading paths. To enable thresholding via a single scalar quantity, we aggregate the per-component nMAEs using the weighted average,
\begin{equation}
    \overline{\mathrm{nMAE}}_e
    =
    \frac
    {\sum\limits_{p=1}^{n_{\mathrm{tar}}}  \omega_p\,\mathrm{nMAE}_p\left(\vct{M}_{i_e^\star}\right)}
    {\sum\limits_{p=1}^{n_{\mathrm{tar}}}  \omega_p},
    \label{eq:inv_nmae_bar}
\end{equation}
where the weights $\omega_p$ can emphasize components that are more critical for the inverse objective.
When all components must individually satisfy prescribed tolerances, componentwise stopping criteria can replace the weighted average. The sequential oracle evaluations are terminated once $\overline{\mathrm{nMAE}}$ falls below a prescribed threshold $\eta$, i.e., $\overline{\mathrm{nMAE}}_e \le \eta$. If the threshold is not met, the oracle evaluations continue until the maximum budget $E_{\max}$ is exhausted.
We denote by $E_\eta$ the oracle evaluation count required to meet the threshold $\eta$:
\begin{equation}
    E_\eta
    =
    \min\left\{e:\ \overline{\mathrm{nMAE}}_e \le \eta\right\},
    \qquad
    E_\eta \le E_{\max}.
    \label{eq:inv_Eeta}
\end{equation}

\paragraph{\textbf{Final selection}} The final design is defined as the best microstructure identified under the prescribed threshold $\eta$ and oracle evaluation budget $E_{\max}$. Specifically,
\begin{equation}
    \vct{M}^\star = \vct{M}_{i^\star},
    \qquad
	    \text{with }
	    i^\star
	    =
	    \begin{cases}
	    i_{E_\eta}^\star, & \text{if } E_\eta \le E_{\max}, \\[1em]
	    \mathop{\arg\min}\limits_{i\in\mathcal{Q}_{E_{\max}}} \overline{\mathrm{nMAE}}_i, & \text{otherwise}
	    \end{cases}.
    \label{eq:inv_final}
\end{equation}
If the threshold $\eta$ is reached within the allotted evaluations, the corresponding design is accepted as a feasible inverse solution. Otherwise, once the oracle evaluation budget $E_{\max}$ is exhausted, the procedure returns the budget-limited optimum, i.e., the candidate within the explored set that yields the smallest aggregated error $\overline{\mathrm{nMAE}}$. This formulation guarantees that the method always returns a valid design: either a design meeting the prescribed threshold or the best attainable solution under the available oracle budget. The complete Bayesian-guided design selection procedure is summarized in \cref{alg:inverse_design}.

\section{Probabilistic surrogate modeling}\label{sec:gp_surrogate}

In our design selection strategy presented in the previous section, we postulated the existence of a probabilistic surrogate that maps microstructures $\vct{M}_i$ to latent variables $\vct{\xi}_i$, which are then transformed via \cref{eq:pos_map} to the material parameters $\vct{\uptheta}_i$ of an effective macroscopic constitutive law \cref{eq:eff_stress}. In this section, we construct such a probabilistic surrogate model in a data-efficient manner. Since microstructures are high-dimensional and oracle calls are costly, we combine dimensionality reduction with Bayesian inference and active learning to achieve accurate predictions with as few labeled samples as possible. The workflow is: 
\begin{enumerate}[label=(\roman*)]
    \item formulate an observation model (\cref{sec:obs_model}),
    \item perform feature engineering to obtain low-dimensional descriptors of microstructures (\cref{sec:feature_engineering}), 
    \item specify a multi-output GP prior over the latent parameters (\cref{sec:multioutput_gp}), 
    \item approximate the posterior using variational Bayesian inference (\cref{sec:vi}), 
    \item derive posterior predictive distributions (\cref{sec:posterior_predictive}), and 
    \item use predictive uncertainty to drive active learning and data acquisition for data-efficient surrogate training (\cref{sec:active_learning}).
\end{enumerate}

\begin{algorithm}[]
\caption{Bayesian-guided design selection}
\label{alg:inverse_design}
\begin{algorithmic}[1]
\setlength{\itemsep}{4pt}
\setlength{\parskip}{0pt}
\setlength{\topsep}{4pt}

\State \textbf{Input:} 
\mbox{designs $\mathcal{D}$},\,
\mbox{deformations $\mathcal{F}$},\,
\mbox{targets $\ten{P}^{\star}$},\,
\mbox{surrogate $\vct{M} \mapsto \vct{\uptheta}$ (\cref{sec:gp_surrogate})},\,
\mbox{budget $E_{\max}$},\,
\mbox{error threshold $\eta$},\,
\mbox{weights $\omega$}

\State Construct shortlist $\mathcal{B}^\star$ via \cref{eq:inv_shortlist}
\State \textbf{Initialize:} $e \gets 1$,\, $\mathcal{Q}_0 \gets \emptyset$,\, $\overline{\mathrm{nMAE}}_0 \gets \infty$

\While{$e \le E_{\max}\ \land\ \overline{\mathrm{nMAE}}_{e-1} > \eta$}
    \State Select next candidate index $i_e^\star$ via \cref{eq:inv_select}
    \State Evaluate oracle response $\ten{P}(\ten{F}_j;\vct{M}_{i_e^\star})$ via \cref{eq:oracle}
    \State Compute aggregated error $\overline{\mathrm{nMAE}}_e$ via \cref{eq:inv_nmae_bar}
    \If{$\overline{\mathrm{nMAE}}_e \le \eta$}
        \State $i^\star \gets i_e^\star$
        \State \textbf{break}
    \EndIf
    \State $e \gets e+1$
\EndWhile

\If{$\overline{\mathrm{nMAE}}_e > \eta$}
    \State $i^\star \gets \mathop{\arg\min}\limits_{i \in \mathcal{Q}_{e}} \overline{\mathrm{nMAE}}_i$
\EndIf

\State \textbf{Output:} $\vct{M}^\star = \vct{M}_{i^\star}$ from \cref{eq:inv_final}

\end{algorithmic}
\end{algorithm}

\subsection{Observation model}
\label{sec:obs_model}
We first specify the observation model from \cref{eq:effective_stress}. In practice, only a subset of components is measured or relevant, and their magnitudes can differ; we therefore apply the operator $\mathcal{S}_{\mathrm{obs}}:\mathbb{R}^{3\times3}\to\mathbb{R}^{n_{\mathrm{obs}}}$, which extracts, scales, and vectorizes the observed stress components used for training, with $n_{\mathrm{obs}}$ the number of such components. The corresponding observations, obtained from the oracle, are collected in the observation vector $\vct{y}^{(i)}_j\in \mathbb{R}^{n_{\mathrm{obs}}}$. Thus, \cref{eq:effective_stress} can be cast as
\begin{equation}
    \vct{y}^{(i)}_j
    =
    \mathcal{S}_{\mathrm{obs}}\Big(
    \widetilde{\ten{P}}\big(\ten{F}_j;\vct{\uptheta}_i(\vct{\xi}_i)\big)
    \Big)
    + \varepsilon^{(i)}_j,
    \label{eq:obs_model}
\end{equation}
where we assume homoscedastic Gaussian noise $\varepsilon^{(i)}_j\in\mathbb{R}^{n_{\mathrm{obs}}}$ with zero mean and variance $\sigma^2$,
\begin{equation}\label{eq:error_model}
    \varepsilon^{(i)}_j \sim \mathcal{N}\left(\vct{0},\, \sigma^2 \ten{I}_{n_{\mathrm{obs}}}\right).
\end{equation}
For a numerical solver, such as the one employed in the present study (\cref{sec:res_Oracle}), repeated evaluations are deterministic when the discretization and solver settings are fixed. The error term therefore represents unresolved effects of solver tolerances, discretization error, and discrepancy arising from the model form. Since the same numerical settings are used consistently for all evaluations and the observed components are scaled to comparable magnitudes, the homoscedastic covariance $\sigma^2\ten{I}_{n_{\mathrm{obs}}}$ provides a parsimonious representation of these aggregate errors. For an experimental oracle, measurement noise, uncertainty in the applied loading and boundary conditions, and specimen-to-specimen variability may instead depend on the observed component and deformation state. These aleatoric effects can be represented by a heteroscedastic covariance, $\varepsilon^{(i)}_j\sim\mathcal{N}(\vct{0},\ten{\Sigma}^{(i)}_j)$, whose identification requires repeated measurements or calibration data spanning the relevant deformation states and experimental conditions.

\subsection{Feature engineering}\label{sec:feature_engineering}
Microstructures are typically represented by high-dimensional pixel/voxel fields (often thousands to millions of variables), making direct surrogate modeling in microstructural image space computationally intractable and statistically redundant. Following established microstructure informatics formulations, we construct a low-dimensional representation using statistical correlation functions followed by PCA-based dimensionality reduction \citep{kalidindi2011microstructure,niezgoda2011understanding,niezgoda2013novel,brough2017materials,kalidindi2020feature}. We express the resulting low-dimensional descriptor through the mapping $\mathcal{A}$,
\begin{equation}
    \vct{z}_i = \mathcal{A}\left(\vct{M}_i\right),
    \qquad
    \mathcal{A}:\mathbb{R}^{n_p}\rightarrow\mathbb{R}^{n_z},
\end{equation}
where $\vct{z}_i\in\mathbb{R}^{n_z}$ denotes the low-dimensional descriptors for the $i$-th design, and $n_z \ll n_p$. In practice, this reduces the representation from thousands/millions of variables to tens or fewer, mitigating the curse of dimensionality while preserving the variability needed for accurate surrogate modeling.
A detailed implementation of this feature engineering workflow is provided in our previous work \citep{danesh2025reduced}; the essential definitions needed here are summarized below.

Let $m_{\vct{g}}^{h,(i)}\in\{0,1\}$ indicate whether the local state $h$ is present at the spatial bin $\vct{g}\in\mathcal{G}$ of the periodic microstructure $\vct{M}_i$. For a spatial shift $\vct{r}$, the corresponding two-point correlation between local states $h$ and $h'$ is
\begin{equation}
    c_{\vct{r}}^{hh',(i)}
    =
    \frac{1}{|\mathcal{G}|}
    \sum_{\vct{g}\in\mathcal{G}}
    m_{\vct{g}}^{h,(i)}
    m_{\vct{g}+\vct{r}}^{h',(i)}.
    \label{eq:feature_correlation}
\end{equation}
The selected correlation fields are variance balanced, vectorized, and concatenated to form the correlation descriptor $\vct{d}_i$. Using the first $n_z$ components, the correlation descriptor is approximated as
\begin{equation}
    \vct{d}_i
    \approx
    \bar{\vct{d}}
    +
    \sum_{k=1}^{n_z} z_{ik}\vct{u}_k,
    \label{eq:feature_pca_representation}
\end{equation}
where $z_{ik}$ is the $k$-th PC score of design $i$, $\vct{u}_k$ is the corresponding PCA basis vector, and $\bar{\vct{d}}$ is the mean correlation descriptor of the ensemble. The correlation statistics and PCA basis are computed once from the unlabeled design set and remain fixed throughout surrogate training and design selection. The particular local states, correlation fields, and retained PCA dimension depend on the application and are specified for a test case in \cref{sec:res_active}. These operations and the final truncated PCA representation collectively define the mapping $\mathcal{A}$.

\subsection{Multi-output Gaussian process prior}\label{sec:multioutput_gp}

With low-dimensional descriptors in hand, we now place a probabilistic surrogate on the latent parameters $\vct{\upxi}_i$. The surrogate maps the reduced features $\vct{z}_i$ to the latent GP outputs $\vct{\upxi}_i$:
\begin{equation}
    \vct{\upxi}_i
    =
    \upxi\left(\vct{z}_i\right),
    \qquad
    \upxi:
    \mathbb{R}^{n_z}\rightarrow\mathbb{R}^{n_\theta}.
\end{equation}
Rather than modeling each of the $n_\theta$ outputs with independent single-output GPs, we construct a multi-output GP that explicitly captures correlation among them. Although the effective stress response \cref{eq:eff_stress} is linear in parameters, the associated energy basis evaluated over the chosen deformation states is generally not composed of mutually independent basis functions. Under the available loading conditions, certain basis functions may be only weakly excited or nearly collinear, such that different parameter combinations can produce nearly indistinguishable stress responses. This limits the identifiability of parameters $\vct{\uptheta}_i$ and induces posterior correlations among them. We therefore adopt a linear model of coregionalization (LMC) \citep{journel1976mining,alvarez2012kernels}, which represents the multi-output covariance as a linear combination of $n_r$ shared latent processes while allowing parameter-specific variability. Each latent process $f_r$ is modeled as an independent GP,
\begin{equation}
    f_r\left(\vct{z}\right) \sim \mathcal{GP}\,\Big(0,\, k_r\big(\vct{z},\vct{z}'\big)\Big),
    \qquad r = 1,\dots,n_r,
\end{equation}
where $k_r(\vct{z},\vct{z}')$ denotes the scalar-valued covariance kernel of the $r$-th latent process over the input space. The latent parameters are then expressed as linear combinations of these latent processes:
\begin{equation}
    \upxi_m\left(\vct{z}\right)
    = \sum_{r=1}^{n_r} a_{mr}\, f_r\left(\vct{z}\right),
    \qquad m = 1,\dots,n_\theta.
\end{equation}
The coefficients $a_{mr}$ determine how strongly each latent process contributes to the $m$-th parameter and thereby encode inter-output correlations. This construction induces a multi-output covariance kernel of LMC form $\ten{\Omega}(\vct{z},\vct{z}')\in\mathbb{R}^{n_\theta\times n_\theta}$, expressed by
\begin{equation}
    \ten{\Omega}\left(\vct{z},\vct{z}'\right)
    =
    \sum_{r=1}^{n_r}
    \vct{a}_r \vct{a}_r^{\top} \, k_r\left(\vct{z},\vct{z}'\right),
    \label{eq:LMC}
\end{equation}
where $\vct{a}_r = ( a_{mr} )_{m=1}^{n_\theta} \in \mathbb{R}^{n_\theta}$ collects the mixing
coefficients associated with the $r$-th latent process. We use the automatic relevance determination squared exponential (ARD-SE) kernel \citep{williams1995gaussian} with $n_z$ length scales, given by
\begin{equation}
    k_r\left(\vct{z},\vct{z}'\right)
    =
    \exp\Big(
    -\frac{1}{2}\sum_{k=1}^{n_z}\frac{\big(z_k-z_k'\big)^2}{\ell_{r,k}^2}
    \Big).
    \label{eq:ard_se_kernel}
\end{equation}
The length scales $\ell_{r,k}$ allow each latent process to weight the low-dimensional descriptors differently and effectively perform automatic relevance determination in the reduced feature space. In summary, we collect the hyperparameters of our multi-output GP in the vector $\vct{\uppsi}_\mathrm{GP}$, which entails the length scales and mixing coefficients:
\begin{equation}
    \vct{\uppsi}_\mathrm{GP}
    =
    \left(
    \{\{\ell_{r,k}\}_{r=1}^{n_r}\}_{k=1}^{n_z},
    \{\vct{a}_r\}_{r=1}^{n_r}
    \right).
    \label{eq:GP_hyper}
\end{equation}
\paragraph{\textbf{Remark}} The ARD-SE kernel encodes the assumption that the latent constitutive parameters vary smoothly with the PC descriptors. This is appropriate for the present fixed microstructure family because the inputs are continuous projections of spatial correlation fields, so nearby descriptors represent statistically similar structures. Moreover, separate length scales for each dimension accommodate unequal sensitivity to the retained PC scores. Automated or generative kernel selection procedures provide alternatives for more heterogeneous data \citep{kuok2025bayesian,abdessalem2017automatic}, while some composite kernels can present identifiability challenges \citep{chen2023identifiability}. We therefore regard the ARD-SE kernel as an empirically adequate choice for the present case rather than a universally optimal kernel.
\subsection{Variational Bayesian inference}\label{sec:vi}

We now formalize the dataset and latent variables for variational Bayesian inference. Let $\ten{Z} = \{\vct{z}_i^\top\}_{i=1}^{N}\in\mathbb{R}^{N\times n_z}$ denote the input matrix for the observed design set, with $N$ the number of observed designs. The corresponding latent parameters are collected in $\vct{\upxi} = \{\vct{\upxi}_i\}_{i=1}^{N}\in\mathbb{R}^{N n_\theta}$, and the physical parameters in $\vct{\uptheta} = \{\vct{\uptheta}_i\}_{i=1}^{N}\in\mathbb{R}^{N n_\theta}$. Let $n_y = n_f n_{\mathrm{obs}}$ denote the total number of stress observations per design and across all deformation states. For each design, we collect all observations into $\vct{y}_i = \{\vct{y}^{(i)}_j\}_{j=1}^{n_f} \in \mathbb{R}^{n_y}$, while $\vct{y} = \{\vct{y}_i\}_{i=1}^{N}\in\mathbb{R}^{N n_y}$ stacks these vectors over all observed designs, yielding the training set $\mathcal{T} = \{(\vct{z}_i,\vct{y}_i)\}_{i=1}^{N}$. 

Given the multi-output LMC kernel $\ten{\Omega}$ from \cref{eq:LMC}, the GP prior over the latent parameters $\vct{\upxi}$ is obtained from
\begin{equation}
    p\left(\vct{\upxi}\mid \ten{Z}\right) = \mathcal{N}\Big(\vct{0},\, \ten{\Omega}\big(\ten{Z},\ten{Z}\big)\Big).
    \label{eq:prior}
\end{equation} 
The latent parameters $\vct{\upxi}$ and the measured data $\vct{y}$ are connected via the observation model \cref{eq:obs_model}, which is used to define the likelihood as
\begin{equation}
    p\left(\vct{y} \mid \vct{\upxi}\right)
    =
    \prod_{i=1}^{N}
    p\left(
    \vct{y}_i \mid \vct{\upxi}_i
    \right),
    \label{eq:likelihood}
\end{equation}
assuming conditional independence of the observations given the latent parameters. Combining the prior \cref{eq:prior} and the likelihood \cref{eq:likelihood} yields the posterior $p(\vct{\upxi}\mid \mathcal{T})$ via Bayes' rule,
\begin{equation}
    p\left(\vct{\upxi}\mid \mathcal{T}\right)
    =
    \frac{p\left(\vct{y}\mid \vct{\upxi}\right)\,p\left(\vct{\upxi}\mid \ten{Z}\right)}
    {p\left(\vct{y}\mid \ten{Z}\right)},
\end{equation}
where $p(\vct{y}\mid \ten{Z})$ is the marginal likelihood (evidence), which normalizes the posterior. The exact posterior is intractable since the positivity map \cref{eq:pos_map} makes the likelihood non-Gaussian, even though the effective constitutive model \cref{eq:eff_stress} is linear in parameters $\vct{\uptheta}_i$. 

To approximate the intractable posterior $p(\vct{\upxi}\mid \mathcal{T})$, we introduce a Gaussian variational distribution $q(\vct{\upxi})$ of the form
\begin{equation}
    q\left(\vct{\upxi}\right) = \mathcal{N}\left(\vct{\upmu},\ten{\Sigma}\right),
    \label{eq:variational_distribution}
\end{equation}
with $\vct{\upmu}\in\mathbb{R}^{N n_\theta}$ the variational mean and $\ten{\Sigma}\in\mathbb{R}^{(N n_\theta)\times(N n_\theta)}$ the variational covariance.
The Gaussian form of the variational distribution is an approximation and does not imply that the exact posterior is Gaussian or provably unimodal. The variational parameters are determined by maximizing the evidence lower bound (ELBO), which follows from Jensen’s inequality \citep{jordan1999introduction,bishop2006pattern}:
\begin{equation}
    \mathcal{L}_{\mathrm{ELBO}}
    =
    \mathbb{E}_{q(\vct{\upxi})}
    \left[
    \log p\left(\vct{y} \mid \vct{\upxi}\right)
    \right]
    -
    \mathrm{KL}\Big(
    q\big(\vct{\upxi}\big)
    \,\|\, 
    p\big(\vct{\upxi}\mid \ten{Z}\big)
    \Big).
    \label{eq:elbo_loss}
\end{equation}
Here, $\mathrm{KL}(\cdot\,\|\,\cdot)$ denotes the Kullback--Leibler divergence, which measures the discrepancy between the variational distribution and the GP prior over the latent parameters, thereby enforcing regularization consistent with the assumed covariance structure. The first term is the expected log-likelihood, which measures agreement with the observed data and is approximated by Monte Carlo integration:
\begin{equation}
    \mathbb{E}_{q(\vct{\upxi})}\left[\log p\left(\vct{y} \mid \vct{\upxi}\right)\right]
    \approx
    \frac{1}{S}\sum_{s=1}^{S}\log p\Big(\vct{y} \mid \vct{\upxi}^{(s)}\Big),
    \qquad
    \vct{\upxi}^{(s)}\sim q\left(\vct{\upxi}\right),
    \label{eq:mc_elbo}
\end{equation}
where $S$ is the total number of samples. Maximizing the $\mathcal{L}_{\mathrm{ELBO}}$ therefore leads to a regularized optimization problem that balances agreement with the observed data and consistency with the GP prior. The corresponding optimal parameter vector is obtained by
\begin{equation}
    \vct{\uppsi}^\star
    =
    \mathop{\arg\max}\limits_{\vct{\uppsi}}
    \mathcal{L}_{\mathrm{ELBO}},
    \label{eq:elbo_opt}
\end{equation}
where $\vct{\uppsi}$ collects all learnable parameters, including the GP hyperparameters $\vct{\uppsi}_\mathrm{GP}$ defined in \cref{eq:GP_hyper}, the observation noise variance $\sigma^2$ appearing in the error model \cref{eq:error_model}, and the variational mean $\vct{\upmu}$ and covariance $\ten{\Sigma}$ introduced in \cref{eq:variational_distribution}:
\begin{equation}
    \vct{\uppsi}
    =
    \left(
    \vct{\uppsi}_\mathrm{GP},\
    \sigma^2,\
    \vct{\upmu},\
    \{\Sigma_{ij}\}_{i,j=1}^{N n_\theta}
    \right).
\end{equation}
This joint optimization yields a consistent probabilistic model whose posterior approximation enables predictive evaluation at new configurations and systematic uncertainty propagation in subsequent computations.

\subsection{Posterior predictive inference}\label{sec:posterior_predictive}
With the variational posterior in hand, we obtain predictive distributions for unseen designs by propagating uncertainty from the latent parameters to stress space. For $N_*$ unseen designs, let $\ten{Z}_* = \{\vct{z}_i^\top\}_{i=1}^{N_*}\in\mathbb{R}^{N_*\times n_z}$ denote the input matrix of low-dimensional descriptors and $\vct{\upxi}_* = \{\vct{\upxi}_i\}_{i=1}^{N_*}\in\mathbb{R}^{N_* n_\theta}$ collect the corresponding latent parameters. For prediction, we marginalize the latent parameters at the observed inputs to obtain the predictive variational distribution,
\begin{equation}
    q\left(\vct{\upxi}_*\right)
    =
    \int
    p\left(\vct{\upxi}_* \mid \vct{\upxi}, \ten{Z}_*, \ten{Z}\right)\, q\left(\vct{\upxi}\right)\, \mathrm{d}\vct{\upxi}.
\end{equation}
Since both the conditional GP prior and the variational posterior are Gaussian, the integral admits a closed form and yields a Gaussian predictive variational distribution $q(\vct{\upxi}_*)=\mathcal{N}(\vct{\upmu}_*,\ten{\Sigma}_*)$. Applying standard Gaussian conditioning identities gives
\begin{equation}
    \vct{\upmu}_* = \ten{\Omega}_{*}^{\top}\ten{\Omega}^{-1}\vct{\upmu},
    \qquad
    \ten{\Sigma}_* = \ten{\Omega}_{**} + \ten{\Omega}_{*}^{\top}\ten{\Omega}^{-1}\left(\ten{\Sigma}-\ten{\Omega}\right)\ten{\Omega}^{-1}\ten{\Omega}_{*},
\end{equation}
which are the closed-form predictive mean $\vct{\upmu}_*\in\mathbb{R}^{N_* n_\theta}$ and covariance $\ten{\Sigma}_*\in\mathbb{R}^{(N_* n_\theta)\times(N_* n_\theta)}$ for the variational GP defined over the observed inputs. Here, $\ten{\Omega}\in\mathbb{R}^{(N n_\theta)\times(N n_\theta)}$, $\ten{\Omega}_{*}\in\mathbb{R}^{(N n_\theta)\times(N_* n_\theta)}$, and $\ten{\Omega}_{**}\in\mathbb{R}^{(N_* n_\theta)\times(N_* n_\theta)}$ denote the covariance matrices evaluated at the observed and unseen inputs:
\begin{equation}
\begin{aligned}
    \ten{\Omega} &= \ten{\Omega}(\ten{Z},\ten{Z}), \\
    \ten{\Omega}_{*} &= \ten{\Omega}(\ten{Z},\ten{Z}_*), \\
    \ten{\Omega}_{**} &= \ten{\Omega}(\ten{Z}_*,\ten{Z}_*).
\end{aligned}
\end{equation}

The uncertainty in the latent parameters $\vct{\upxi}_*$ is then propagated to the effective material parameters $\vct{\uptheta}_* = \{\vct{\uptheta}_i\}_{i=1}^{N_*}\in\mathbb{R}^{N_* n_\theta}$ via the positivity map \cref{eq:pos_map}. The forecasted stresses $\widetilde{\vct{y}}_*\in
    \mathbb{R}^{N_* n_y}$ follow accordingly from the macroscopic effective model \cref{eq:eff_stress}:
\begin{equation}
    \widetilde{\vct{y}}_*
    =
    \left\{
    \left\{
    \mathcal{S}_{\mathrm{obs}}\Big(
    \widetilde{\ten{P}}\big(\ten{F}_j;\vct{\uptheta}_i\big)
    \Big)
    \right\}_{j=1}^{n_f}
    \right\}_{i=1}^{N_*}.
    \label{eq:stress_pushforward}
\end{equation}
Given $S$ Monte Carlo samples, the predictive moments are then estimated by
\begin{equation}
    \mathbb{E}\left[\widetilde{\vct{y}}_*\right] \approx \frac{1}{S}\sum_{s=1}^{S}\widetilde{\vct{y}}_*^{(s)},
    \qquad
    \mathrm{Var}\left[\widetilde{\vct{y}}_*\right] \approx \frac{1}{S-1}\sum_{s=1}^{S}\Big(\widetilde{\vct{y}}_*^{(s)}-\mathbb{E}\left[\widetilde{\vct{y}}_*\right]\Big)\odot\Big(\widetilde{\vct{y}}_*^{(s)}-\mathbb{E}\left[\widetilde{\vct{y}}_*\right]\Big),
    \label{eq:mc_obs}
\end{equation}
where $\odot$ denotes componentwise multiplication. The samples $\widetilde{\vct{y}}_*^{(s)}$ correspond to draws $\vct{\upxi}_*^{(s)} \sim q(\vct{\upxi}_*)$, mapped to $\vct{\uptheta}_*^{(s)}$ via the positivity transformation \cref{eq:pos_map}, from which the stress predictions are computed according to \cref{eq:stress_pushforward}. This formulation ensures consistent propagation of latent uncertainty onto the predicted stress response and provides the probabilistic foundation for uncertainty-driven model exploration (\cref{sec:active_learning}) and uncertainty-aware design decisions (\cref{sec:inv_des}).

\subsection{Uncertainty-driven active learning}
\label{sec:active_learning}
With the predictive stress distributions available from the previous section, we now leverage uncertainty to guide data acquisition through an iterative active learning procedure \citep{liu2024active,buzzy2025active,ozbayram2025batch,danesh2025reduced}. The goal is to train the surrogate with as few oracle evaluations as possible by selecting, at each iteration, the designs expected to provide the greatest information gain. To formalize the iterative procedure, at iteration $t\in\{0,\dots,T_{\max}\}$, we partition the design index set $\mathcal{I}$ into three subsets: the labeled set $\mathcal{I}^{\mathrm{lab}}_t$, the unlabeled set $\mathcal{I}^{\mathrm{unlab}}_t$, and the test set $\mathcal{I}^{\mathrm{test}}$ such that
\begin{equation}
\begin{aligned}
    \mathcal{I}^{\mathrm{lab}}_t \cup \mathcal{I}^{\mathrm{unlab}}_t \cup \mathcal{I}^{\mathrm{test}} &= \mathcal{I},\\
    \mathcal{I}^{\mathrm{lab}}_t \cap \mathcal{I}^{\mathrm{unlab}}_t &= \emptyset,\\
    \mathcal{I}^{\mathrm{lab}}_t \cap \mathcal{I}^{\mathrm{test}} &= \emptyset,\\
    \mathcal{I}^{\mathrm{unlab}}_t \cap \mathcal{I}^{\mathrm{test}} &= \emptyset, \qquad \forall\, t.
\end{aligned}
\end{equation}
The labeled set provides training data, the unlabeled pool provides candidates for acquisition, and the test set is held out to monitor predictive performance. We denote the current labeled training set, unlabeled pool, and fixed hold-out set by
\begin{equation}
    \mathcal{T}_t = \left\{\left(\vct{z}_i,\vct{y}_i\right)\right\}_{i\in\mathcal{I}^{\mathrm{lab}}_t},
    \qquad
    \mathcal{P}_t = \left\{\vct{z}_i\right\}_{i\in\mathcal{I}^{\mathrm{unlab}}_t},
    \qquad
    \mathcal{H} = \left\{\left(\vct{z}_i,\vct{y}_i\right)\right\}_{i\in\mathcal{I}^{\mathrm{test}}}.
\end{equation}

At iteration $t=0$, the initial index sets $\mathcal{I}^{\mathrm{lab}}_0$, $\mathcal{I}^{\mathrm{unlab}}_0$, and $\mathcal{I}^{\mathrm{test}}$ (and thus $\mathcal{T}_0$ and $\mathcal{P}_0$) are formed by a chosen sampling strategy over $\mathcal{D}$, such as random sampling, Latin hypercube sampling, clustering-based selection, or other space-filling designs, to ensure a representative starting coverage of the feasible set. The initial labeled set is then populated by oracle evaluations for each $i\in\mathcal{I}^{\mathrm{lab}}_0$, which yields the observed stresses $\vct{y}_{i}\in\mathbb{R}^{n_y}$ needed to construct $\mathcal{T}_0$ according to
\begin{equation}
    \vct{y}_{i}
    =
    \mathcal{S}_{\mathrm{obs}}\Big(
    \left\{\mathcal{O}\big(\ten{F}_j;\vct{M}_{i}\big)\right\}_{j=1}^{n_f}
    \Big).
    \label{eq:al_oracle}
\end{equation}
The surrogate is then trained on $\mathcal{T}_0$ by solving the ELBO optimization problem \cref{eq:elbo_opt}, providing the initial predictive model used for acquisition.

\begin{figure}
    \centering
    \includegraphics{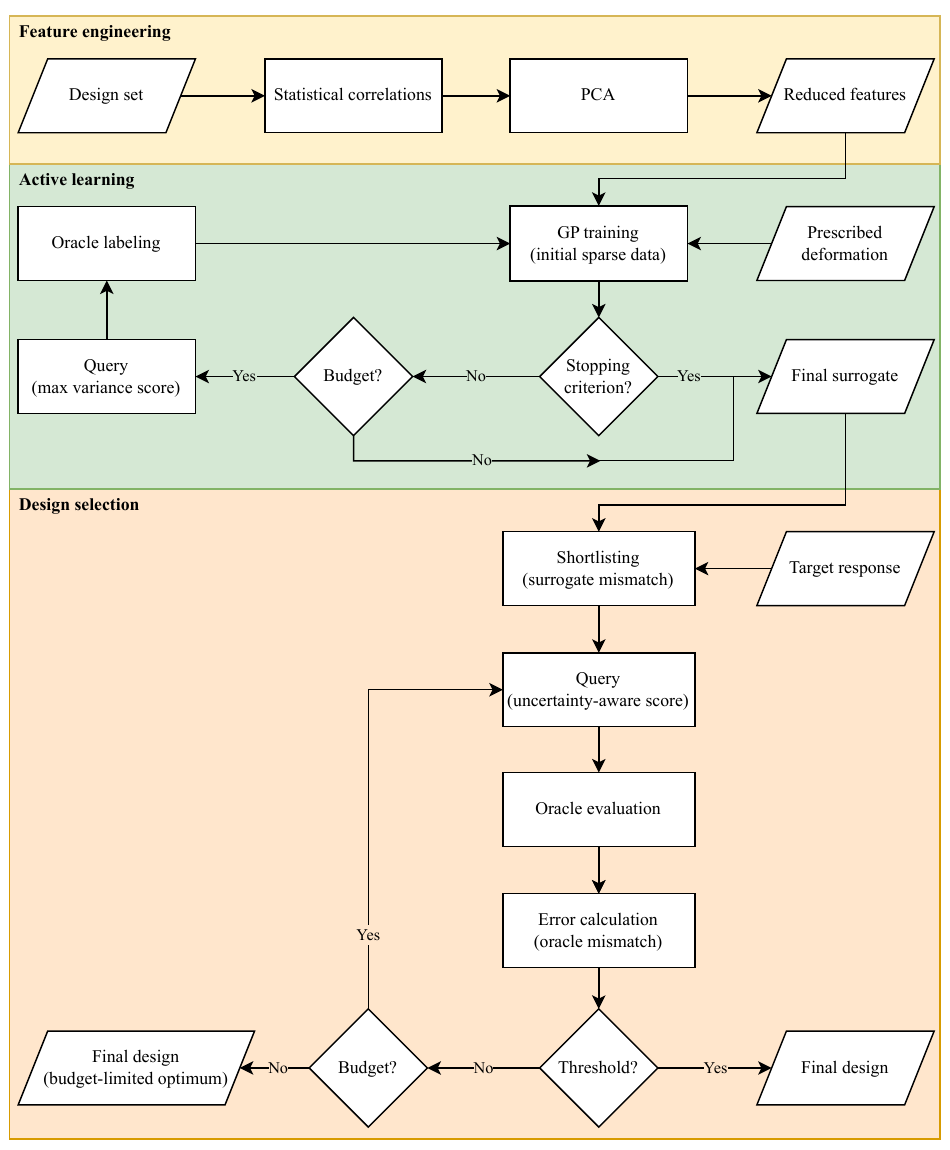}
    \caption{Flowchart of the methodological framework: feature engineering produces reduced microstructural descriptors, uncertainty-driven active learning trains the GP surrogate with minimal labeled data, and Bayesian-guided design selection uses the trained model to identify a feasible design while minimizing oracle evaluations.}
    \label{fig:flow}
\end{figure}

At iteration $t>0$, we evaluate an acquisition score for each candidate in the unlabeled pool $\mathcal{P}_{t-1}$ by quantifying the predictive uncertainty of its stress response,
\begin{equation}
    \alpha_t\left(\vct{z}_i\right)
    =
    \log \operatorname{det}\Big(
    \operatorname{diag}\big(
    \mathrm{Var}(\widetilde{\vct{y}}_* \mid \mathcal{T}_{t-1})
    \big)_{(i)}
    \Big),
    \qquad
    \forall\, i \in \mathcal{I}^{\mathrm{unlab}}_{t-1}.
    \label{eq:al_acq}
\end{equation}
This acquisition criterion assigns higher scores to candidates exhibiting large predictive variance across stress components. This criterion has proven effective in practice \citep{liu2024active,buzzy2025active,ozbayram2025batch,danesh2025reduced}, particularly when the model noise is approximately uniform across the input space \cite{settles2009active}. Since the predictive stress distribution is obtained as the pushforward of the variational approximation (cf. \cref{eq:stress_pushforward}), the variance reflects epistemic uncertainty induced by limited training data. Maximizing the acquisition function therefore promotes sampling in regions where the current model exhibits the largest uncertainty in the predicted stress response, thereby targeting maximal expected information gain. At iteration $t$, the next configuration is selected by
\begin{equation}
    i_t^\star = \mathop{\arg\max}\limits_{i\in\mathcal{I}^{\mathrm{unlab}}_{t-1}} \alpha_t\left(\vct{z}_i\right),
    \label{eq:al_select}
\end{equation}
and then moved from the unlabeled set to the labeled set, yielding the updated index sets
\begin{equation}
    \mathcal{I}^{\mathrm{lab}}_{t} = \mathcal{I}^{\mathrm{lab}}_{t-1} \cup \{i_t^\star\},
    \qquad
    \mathcal{I}^{\mathrm{unlab}}_{t} = \mathcal{I}^{\mathrm{unlab}}_{t-1} \setminus \{i_t^\star\}.
    \label{eq:al_index_update}
\end{equation}
The oracle is evaluated for the selected design $\vct{M}_{i_t^\star}$ to obtain new stress observations using \cref{eq:al_oracle}. The resulting observations are then appended to the training set while the queried point is removed from the pool,
\begin{equation}
    \mathcal{T}_{t}
    =
    \mathcal{T}_{t-1} \cup \left\{\left(\vct{z}_{i_t^\star},\vct{y}_{i_t^\star}\right)\right\},
    \qquad
    \mathcal{P}_{t}
    =
    \mathcal{P}_{t-1} \setminus \{\vct{z}_{i_t^\star}\}.
    \label{eq:al_update}
\end{equation}
The surrogate model is then re-trained using the augmented training set $\mathcal{T}_{t}$ by re-optimizing the ELBO objective \cref{eq:elbo_opt}, while predictive accuracy is monitored on a fixed hold-out set via
\begin{equation}
    \mathrm{MAE}_{t}
    =
    \frac{1}{|\mathcal{H}|}
    \sum_{i\in\mathcal{I}^{\mathrm{test}}}
    \left\|
    \vct{y}_i
    -
    \left(
    \mathbb{E}\left[\widetilde{\vct{y}}_* \mid \mathcal{T}_t\right]
    \right)_{(i)}
    \right\|_{1},
    \qquad
    \forall\,i\in\mathcal{I}^{\mathrm{test}}.
    \label{eq:al_mae}
\end{equation}
The MAE evaluated on the hold-out set provides a consistent out-of-sample measure of predictive accuracy as the surrogate model is iteratively re-trained. The stopping criterion is defined in terms of the relative change in MAE computed over a sliding window \citep{danesh2025reduced},
\begin{equation}
    \Delta_t
    =
    \frac{1}{L}
    \sum_{t=T-L+1}^{T}
    \frac{\big|
    \mathrm{MAE}_{t}
    -
    \mathrm{MAE}_{t-1}
    \big|}
    {\mathrm{MAE}_{t-1}},
    \label{eq:al_stop}
\end{equation}
where $L$ is the window length and $T$ denotes the current iteration index used to compute the sliding average. The active learning procedure is terminated when improvements fall below a prescribed threshold, $\Delta_t \leq \epsilon$, or when the oracle evaluation budget is exhausted (i.e., $t \geq T_{\max}$). This rule prevents additional oracle evaluations once the surrogate exhibits negligible improvement in predictive accuracy. \cref{alg:active_learning} formalizes the  complete active learning procedure, and \cref{fig:flow} illustrates its integration within the overall methodological framework.

\begin{algorithm}[]
\caption{Active learning procedure}
\label{alg:active_learning}
\begin{algorithmic}[1]
\setlength{\itemsep}{4pt}
\setlength{\parskip}{0pt}
\setlength{\topsep}{4pt}

\State \textbf{Input:}
\mbox{designs $\mathcal{D}$},\,
\mbox{deformations $\mathcal{F}$},\,
\mbox{design indices $\mathcal{I}$},\,
\mbox{labeled indices $\mathcal{I}^{\mathrm{lab}}_0$},\,
\mbox{unlabeled indices $\mathcal{I}^{\mathrm{unlab}}_0$},\,
\mbox{test indices $\mathcal{I}^{\mathrm{test}}$},\,
\mbox{training set $\mathcal{T}_0$},\,
\mbox{pool $\mathcal{P}_0$},\,
\mbox{test set $\mathcal{H}$},\,
\mbox{budget $T_{\max}$},\,
\mbox{averaging window $L$},\,
\mbox{stopping threshold $\epsilon$}

\State \textbf{Initialize:} $t \gets 0$,\, $\Delta_0 \gets \infty$

\State Train surrogate on $\mathcal{T}_0$ by solving \cref{eq:elbo_opt} to obtain $\vct{\uppsi}_0^\star$

\While{$t < T_{\max}\ \land\ \Delta_t \ge \epsilon$}
    \State $t \gets t + 1$
    \State Compute acquisition scores $\alpha_{t}(\vct{z}_i)$ for all $i \in \mathcal{I}^{\mathrm{unlab}}_{t-1}$ via \cref{eq:al_acq}
    \State Select next query $i_t^\star$ via \cref{eq:al_select}
    \State Query oracle to obtain $\vct{y}_{i_t^\star}$ via \cref{eq:al_oracle}
    \State Update index sets $\mathcal{I}^{\mathrm{lab}}_{t}$ and  $\mathcal{I}^{\mathrm{unlab}}_{t}$ via \cref{eq:al_index_update}
    \State Update training set $\mathcal{T}_{t}$ and pool $\mathcal{P}_{t}$ via \cref{eq:al_update}
    \State Re-train surrogate on $\mathcal{T}_{t}$ by re-optimizing \cref{eq:elbo_opt} to obtain $\vct{\uppsi}_{t}^\star$
    \State Evaluate $\mathrm{MAE}_t$ on $\mathcal{H}$ via \cref{eq:al_mae} and update $\Delta_t$ via \cref{eq:al_stop}
\EndWhile

\State \textbf{Output:} $\vct{\uppsi}_{t}^\star$ (final surrogate model)
\end{algorithmic}
\end{algorithm}

\section{Results and discussion}

To assess the performance of the proposed framework, we apply it to a large and diverse design set of stochastic hyperelastic metamaterials. We first introduce the discrete design set (\cref{sec:res_Des}) and the high-fidelity oracle used to compute the effective homogenized response (\cref{sec:res_Oracle}). Subsequently, we define a physics-consistent effective constitutive model that approximates the high-fidelity solutions (\cref{sec:res_eff}) and establish an appropriate sampling strategy in deformation gradient space (\cref{sec:res_sampling}). Based on these ingredients, we train a data-efficient surrogate model via active learning to infer the effective constitutive parameters from limited oracle evaluations in \cref{sec:res_active}. Finally, the trained surrogate is employed to perform Bayesian-guided design selection over the considered set of stochastic metamaterial unit cells (\cref{sec:res_inverse}).

\subsection{Design set: stochastic metamaterial unit cells}
\label{sec:res_Des}

We employ the design set of periodic two-dimensional stochastic metamaterials introduced by \citet{bastek2023inverse,bastek2023data}, comprising $53,019$ unit cells represented on a $96 \times 96$ pixel grid. Each unit cell is denoted by $\vct{M}_i \in \mathbb{R}^{n_p}$, where $\vct{M}_i$ is the vectorized representation of the pixel grid and $n_p=96\times 96=9,216$. The original design set \citep{bastek2023inverse,bastek2023data} intentionally includes geometries exhibiting instability modes in order to study buckling behavior. In the present work, we exclude configurations identified as buckling-prone, resulting in the design set $\mathcal{D}=\{\vct{M}_i\}_{i=1}^{n_d}$ with $n_d=50,000$ admissible unit cells. In our previous work \citep{danesh2025reduced,danesh2025data}, this design set served as the foundation for developing reduced-order structure--property linkages for effective elastic constants. 

Each microstructure is generated by sampling a Gaussian random field followed by threshold-based binarization to obtain a solid--void representation. Admissible samples are mirrored to enforce periodicity, resulting in a geometrically diverse collection of stochastic unit cells (see \citep{bastek2023inverse} for more details). Notably, this microstructure generator is non-differentiable, but enables the computationally efficient generation of a large and diverse design set. 

\begin{figure}[h]
    \centering
    \includegraphics[width=0.5\linewidth]{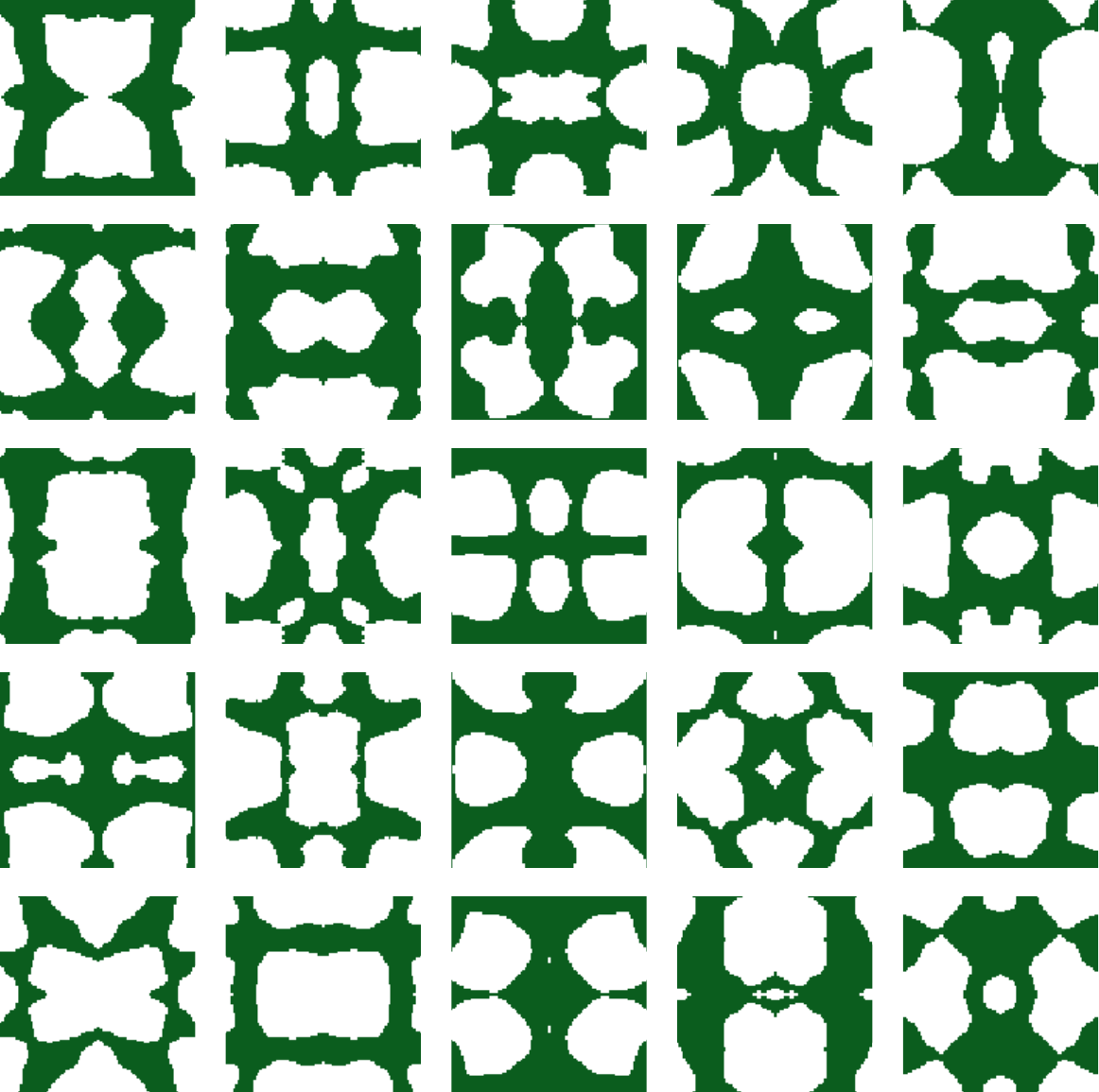}
    \caption{Representative subset of 25 randomly selected stochastic metamaterial unit cells from a design set of 50,000 candidates, highlighting the diversity of the generated designs.}
    \label{fig:sample_designs}
\end{figure}

 The geometric variability of this design set was quantitatively assessed in our previous study \citep{danesh2025reduced}, where diversity metrics were reported to characterize coverage of the design space. The generated microstructures span a broad range of relative densities, from 0.30 to 0.68, and exhibit a high degree of morphological variability, with a normalized average pairwise Euclidean distance of 0.69 between binary structures. Furthermore, k-means clustering with 500 clusters produced a low coefficient of variation in cluster sizes (0.19), indicating that the design set is broadly distributed throughout the design space rather than concentrated in a limited set of morphologies. Such broad geometric coverage is particularly important here, as the proposed design selection strategy operates by selection over a finite candidate library rather than continuous optimization. Ensuring sufficient diversity within the discrete pool increases the likelihood that the prescribed target response can be approximated within the available design space. A representative subset of randomly selected unit cells is shown in \cref{fig:sample_designs}, highlighting the diversity of the generated designs.

\subsection{Oracle: FFT-based computational homogenization}
\label{sec:res_Oracle}

To obtain homogenized effective responses of the stochastic unit cells, a high-fidelity oracle is required. In principle, such an oracle may consist of experimental measurements or numerical simulations. In the present study, we employ computational homogenization to compute the effective stress response of each unit cell under prescribed deformation states and periodic boundary conditions. The solid phase is modeled as an incompressible Neo-Hookean material under a plane-stress formulation appropriate for thin metamaterial films. The shear modulus of the solid phase is set to 100 MPa, whereas the void phase is modeled as a compliant material with stiffness 1 MPa (contrast ratio of 0.01), consistent with the contrast employed for spinodoid architected structures in \citep{rassloff2025inverse, otto2025data}. This moderate contrast mitigates convergence issues associated with extreme material contrasts in spectral solvers \cite{lucarini2022adaptation,danesh2023challenges}.

Homogenization is performed using the Galerkin FFT method \citep{vondvrejc2014fft,vondvrejc2015guaranteed}, combined with a rotated finite difference discretization \citep{willot2015fourier} and solved using the MINRES iterative solver, a combination suggested by \citet{lucarini2022adaptation} to adapt spectral solvers for metamaterial unit cells. In such FFT-based formulations, the periodic boundary conditions are inherently satisfied as the underlying fields are represented periodically. This framework has been employed in our previous work to generate high-fidelity datasets, including homogenized elastic constants of auxetic unit cells \citep{danesh2024fft} and the present stochastic design set \citep{danesh2025reduced} in the small-strain regime. While those earlier studies focused on small-strain elasticity, the present work employs a finite-strain formulation \citep{de2017finite,minh2020surrogate} adapted to plane-stress assumption to capture the nonlinear response of thin hyperelastic metamaterial films. In this setting, the oracle evaluations defined in \cref{eq:oracle} correspond to the effective first Piola stress $\ten{P}(\ten{F}_j;\vct{M}_i)$ obtained from representative volume element (RVE) simulations of the unit cell geometry $\vct{M}_i$ under prescribed macroscopic deformation gradients $\ten{F}_j$. In what follows, the dependence on the particular RVE and loading state is omitted for notational simplicity.

To briefly summarize the homogenization procedure, let $\vct{X}\in V_0\subset\mathbb{R}^3$ denote the material coordinate in the reference configuration of an RVE (i.e., a metamaterial unit cell in our case), with $V_0$ the reference domain. The microscopic deformation gradient $\ten{F}^{\mu}$ is decomposed into a prescribed macroscopic part $\ten{F}$ and a fluctuation field
$\ten{F}'(\vct{X})$,
\begin{equation}
    \ten{F}^{\mu}\left(\vct{X}\right)
    =
    \ten{F}
    +
    \ten{F}'\left(\vct{X}\right),
\end{equation}
where the fluctuation field $\ten{F}'(\vct{X})$ is periodic over $V_0$ and has vanishing volume average over the RVE. Accordingly, the macroscopic deformation gradient equals
the volume average of the microscopic field,
\begin{equation}
    \ten{F}
    =
    \frac{1}{|V_0|}
    \int_{V_0}
    \ten{F}^{\mu}\left(\vct{X}\right)\,\mathrm{d}V.
\end{equation}
Analogously, the effective first Piola stress $\ten{P}$ is obtained by volume
averaging the microscopic stress field $\ten{P}^{\mu}$:
\begin{equation}
    \ten{P}
    =
    \frac{1}{|V_0|}
    \int_{V_0}
    \ten{P}^{\mu}\left(\vct{X}\right)\,\mathrm{d}V.
    \label{eq:avg_P}
\end{equation}

Importantly, the proposed design selection framework is independent of the specific homogenization scheme. The Galerkin FFT solver serves solely as a provider of high-fidelity response data. In principle, the oracle could be replaced by alternative numerical methods or experimental measurements without altering the methodology. Detailed aspects of the Galerkin FFT formulation are therefore beyond the scope of the present work, and interested readers are referred to \citep{vondvrejc2014fft,vondvrejc2015guaranteed,willot2015fourier,danesh2024fft,lucarini2022adaptation,de2017finite,minh2020surrogate} for further discussion.

\subsection{Effective macroscopic constitutive model}
\label{sec:res_eff}

The unit cell generation strategy, including random construction combined with periodic mirroring (cf.~\cref{sec:res_Des}), induces an orthotropic effective response with orthogonal in-plane material axes. In particular, each generated cell is symmetric by construction with respect to reflections about these two axes, which implies that the homogenized behavior is direction-dependent and invariant under sign reversal (reflection) along each preferred axis. Following \citep{holzapfel2000new,holzapfel2002nonlinear, mcculloch2024automated}, we denote the corresponding preferred directions by two orthogonal unit vectors $\vct{a}_0\in\mathbb{R}^3$ and $\vct{b}_0\in\mathbb{R}^3$, i.e.,
\begin{equation}
    \|\vct{a}_0\|_2 = 1, 
    \qquad 
    \|\vct{b}_0\|_2 = 1, 
    \qquad 
    \vct{a}_0 \cdot \vct{b}_0 = 0,
\end{equation}
which define an orthotropic symmetry class. In the present setting, these preferred directions coincide with the axes of the reference coordinate system due to the symmetric construction of the unit cells, such that $\vct{a}_0 = \vct{e}_1$ and $\vct{b}_0 = \vct{e}_2$. To represent this symmetry in an objective manner, we introduce the invariant basis
\begin{equation}
    I_1 = \operatorname{tr}\left(\ten{F}^{\top} \ten{F}\right),
    \qquad
    I_4
    =
    \vct{a}_0 \cdot \left(\ten{F}^{\top}\ten{F}\,\vct{a}_0\right)
    =
    \|\ten{F}\,\vct{a}_0\|_2^2,
    \qquad
    I_6
    =
    \vct{b}_0 \cdot \left(\ten{F}^{\top}\ten{F}\,\vct{b}_0\right)
    =
    \|\ten{F}\,\vct{b}_0\|_2^2.
\end{equation}
Here, $I_1$ captures the isotropic distortion, while $I_4$ and $I_6$ represent the squared stretches along the two preferred material directions.

Based on this orthotropic invariant basis, we define the effective constitutive model \cref{eq:eff_stress}. For readability, we omit the design index $i$ and deformation state index $j$ and consider a representative structure under a generic deformation state. In such an anisotropic case, the effective strain energy depends not only on the deformation gradient but also on the preferred directions $\vct{a}_0$ and $\vct{b}_0$, i.e.,
\begin{equation}
    \widetilde{W}\left(\ten{F}, \vct{a}_0, \vct{b}_0;\vct{\uptheta}\right)
    =
    \widetilde{W}\left(I_1, I_4, I_6;\vct{\uptheta}\right).
\end{equation}
Since the preferred directions $\vct{a}_0$ and $\vct{b}_0$ are fixed by the symmetry of the unit cell, their explicit appearance will be omitted in the following notation for brevity. Specifying \cref{eq:energy_linear}, we introduce a compact energy ansatz as a linear combination of invariant basis terms plus an incompressibility constraint,
\begin{equation}
    \widetilde{W}\left(\ten{F};\vct{\uptheta}\right)
    =
    \vct{Q}\left(I_1, I_4, I_6\right)\cdot \vct{\uptheta}
    -
    p\,\left(J-1\right),
    \qquad
    \vct{Q}\left(I_1, I_4, I_6\right)
    =
    \begin{bmatrix}
    \left(I_1-3\right)\\
    \left(I_4-1\right)^2\\
    \left(I_6-1\right)^2
    \end{bmatrix},
    \quad
    \vct{\uptheta}
    =
    \begin{bmatrix}
    \theta_{I_1}\\
    \theta_{I_4}\\
    \theta_{I_6}
    \end{bmatrix},
\end{equation}
which can be written in expanded invariant form:
\begin{equation}
    \widetilde{W}\left(\ten{F};\vct{\uptheta}\right)
    =
    \theta_{I_1}\,\left(I_1-3\right)
    +
    \theta_{I_4}\,\left(I_4-1\right)^2
    +
    \theta_{I_6}\,\left(I_6-1\right)^2
    -
    p\,\left(J-1\right).
    \label{eq:Wtilde_invariant_form}
\end{equation}
The effective first Piola--Kirchhoff stress is then obtained from \cref{eq:eff_stress} as
\begin{equation}
    \widetilde{\ten{P}}\left(\ten{F};\vct{\uptheta}\right)
    =
    2\,\theta_{I_1}\,\ten{F}
    +
    4\,\theta_{I_4}\,\left(I_4-1\right)\,\ten{F}\,\left(\vct{a}_0\otimes\vct{a}_0\right)
    +
    4\,\theta_{I_6}\,\left(I_6-1\right)\,\ten{F}\,\left(\vct{b}_0\otimes\vct{b}_0\right)
    -
    p\,\ten{F}^{-\top}.
    \label{eq:pk1_theta_form}
    \end{equation}
Accordingly, to enforce incompressible plane-stress conditions, the hydrostatic pressure $p$ is determined from the out-of-plane stress condition $P_{33}=0$, yielding
    \begin{equation}
    p = 2\,\theta_{I_1}\,\frac{F_{33}}{\operatorname{cof}\left(\ten{F}\right)_{33}}.
    \label{eq:p_cof_form_incomp}
\end{equation}

\paragraph{\textbf{Remark}} In this representation, the linear contribution in $I_1$, with parameter $\theta_{I_1}$, represents the isotropic part of the effective response, consistent with the Neo-Hookean constitutive law used for the solid phase in the high-fidelity oracle (cf.~\cref{sec:res_Oracle}). The additional quadratic terms in $I_4$ and $I_6$, with parameters $\theta_{I_4}$ and $\theta_{I_6}$, respectively, introduce directional stiffening along the two preferred material directions while preserving zero stress in the reference configuration. In this sense, the model can be interpreted as a minimal orthotropic extension of the underlying Neo-Hookean solid phase behavior to the homogenized unit cell response. Furthermore, one may additionally include a mixed interaction invariant of the form $\vct{a}_0 \cdot (\ten{F}^{\top}\ten{F}\,\vct{b}_0)$. In the present work, however, we restrict the effective basis to $\{ I_1, I_4, I_6 \}$, since the axis-aligned deformation states introduced below render this interaction term identically zero. An analogous behavior is reported in orthotropic invariant formulations of textile structures under aligned biaxial loading \citep{mcculloch2024automated}, where the interaction invariant vanishes for axis-aligned stretch states.

\subsection{Deformation sampling}
\label{sec:res_sampling}

We consider two different deformation sampling strategies: one used to generate the \emph{training data} for the surrogate model and another used to construct the \emph{target data} to evaluate the performance of our design selection strategy. This distinction separates the surrogate training phase from the assessment of the proposed design selection methodology. The surrogate is trained on a structured set of deformation states aligned with the preferred material directions of the unit cell, enabling efficient identification of the constitutive response. In contrast, target responses are generated from deformation states that differ from those encountered during training, allowing us to systematically assess the robustness of the inferred constitutive model and the design selection framework under off-axis loading configurations not explicitly included during training.

\paragraph{\textbf{Training data}} To identify the effective parameters reliably, the sampled deformation states must sufficiently excite the invariant contributions associated with $I_1$, $I_4$, and $I_6$. We therefore define a biaxial in-plane loading family analogous to that introduced in \citep{mcculloch2024automated}, consisting of the five paths listed in \cref{tab:loading_paths}: Tension-x, Off-x, Equibiaxial, Off-y, and Tension-y. These paths vary the principal stretches in a structured manner, ranging from pure tension in one preferred direction to unbalanced and balanced biaxial tension, ensuring systematic excitation of the invariant contributions. Since the metamaterial thin films considered here exhibit out-of-plane instabilities under compressive or shear-dominated in-plane loading, the deformation sampling is restricted to tensile states.

\begin{table}[h]
    \centering
    \caption{Loading paths and maximum principal stretches used for deformation sampling.}
    \label{tab:loading_paths}
    \begin{tabular}{lcc}
        \hline
        Loading path & $\lambda_1^{\max}$ & $\lambda_2^{\max}$ \\
        \hline
        Tension-x    & 1.50 & 1.00 \\
        Off-x        & 1.50 & 1.25 \\
        Equibiaxial  & 1.50 & 1.50 \\
        Off-y        & 1.25 & 1.50 \\
        Tension-y    & 1.00 & 1.50 \\
        \hline
    \end{tabular}
\end{table}

Under the adopted biaxial tension loading, we consider incompressible diagonal in-plane deformations aligned with the preferred material directions, whose deformation gradient is given by
\begin{equation}
    \ten{F}
    =
    \mathrm{diag}\Big(
    \lambda_1,\,
    \lambda_2,\,
    \big(\lambda_1 \lambda_2\big)^{-1}
    \Big),
    \label{eq:diagonal_F}
\end{equation}
where $\lambda_1$ and $\lambda_2$ denote the principal stretches along the first and second preferred directions, respectively. This construction enforces incompressibility, satisfies the plane-stress assumption, and preserves alignment between the deformation axes and the orthotropic material directions. For such a class of deformations, the invariants simplify to
\begin{equation}
    I_1
    =
    \lambda_1^2
    +
    \lambda_2^2
    +
    \left(\lambda_1 \lambda_2\right)^{-2},
    \qquad
    I_4 = \lambda_1^2,
    \qquad
    I_6 = \lambda_2^2.
\end{equation}
Accordingly, $I_4$ and $I_6$ are directly controlled by the principal stretches, while $I_1$ captures the coupled distortion induced by incompressibility.

\begin{figure}[t]
    \centering
    \includegraphics[width=\linewidth]{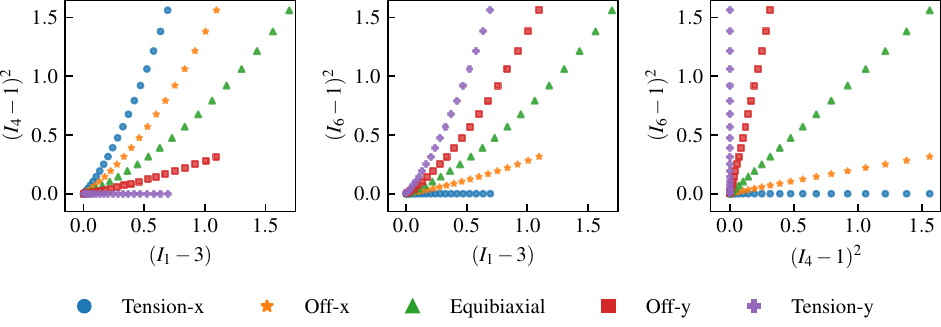}
    \caption{Sampling of the five loading paths in invariant space. The panels show pairwise relationships between $(I_1-3)$, $(I_4-1)^2$, and $(I_6-1)^2$ for the sampled deformation states obtained from the incompressible plane-stress deformation gradient.}
    \label{fig:deformation_sampling}
\end{figure}

For each loading path listed in \cref{tab:loading_paths}, the principal stretches are discretized as
\begin{equation}\label{eq:lambda_min_max}
    \lambda_k \in 
    \left\{
    1 + \frac{h}{n_\lambda}\left(\lambda_k^{\max}-1\right)
    \right\}_{h=0}^{n_\lambda},
    \qquad k = 1,2,
\end{equation}
with $n_\lambda = 20$ deformation increments per path. This structured sampling generates distinct trajectories in invariant space, which are visualized in \cref{fig:deformation_sampling}. The separation of these trajectories promotes independent variation of $(I_1-3)$, $(I_4-1)^2$, and $(I_6-1)^2$, thereby reducing parameter collinearity and improving identifiability of the effective orthotropic coefficients.

Under the considered biaxial loading configuration aligned with the preferred material directions (\cref{sec:res_eff}), the imposed deformation gradient is diagonal in the material frame (i.e., \cref{eq:diagonal_F}) and contains no shear components. Owing to the alignment of the loading axes with the orthotropic symmetry directions of the microstructure, the overall response remains coaxial with the imposed deformation. Consequently, the homogenized first Piola--Kirchhoff stress tensor is diagonal in the material frame, such that the only nonzero in-plane components are $P_{11}$ and $P_{22}$, while all shear components vanish by symmetry. The out-of-plane stress component is also eliminated by enforcing the plane-stress condition $P_{33}=0$. In experimental settings, the corresponding nominal stresses $P_{11}$ and $P_{22}$ are computed from the measured reaction forces along the loading axes together with the associated reference cross-sectional areas. In the present work, the same quantities are obtained from physics-based computational homogenization (cf. \cref{sec:res_Oracle}). These nominal stresses $P_{11}$ and $P_{22}$ are then used as the stress observations to train the surrogate.

\paragraph{\textbf{Target data}} The surrogate is trained using loading paths aligned with the preferred material directions of the unit cell. As target data for the proposed design selection framework, we deliberately consider off-axis loading configurations that were not explicitly encountered during surrogate training. In experimental settings, rotating a specimen with respect to the applied loading direction is conceptually straightforward. In numerical simulations, however, such a strategy is not feasible, since rotating the discretized microstructure would interfere with the imposed periodic boundary conditions and alter the underlying grid representation. Instead, we mimic off-axis loading by rotating the deformation gradient while keeping the microstructure fixed. This operation does not correspond to a superposed rigid body rotation, but rather to a re-expression of the prescribed stretch in a rotated loading coordinate system. This preserves the material frame associated with the unit cell and provides a clean and controlled assessment of the surrogate response under an off-axis loading orientation.

To formalize this idea, we introduce a rotation of the loading frame by an angle $\beta$ about the out-of-plane axis $\vct{e}_3$. Let $\ten{R}$ denote the corresponding rotation tensor. The rotated deformation gradient is then obtained by a similarity transformation of the original deformation gradient $\ten{F}$, i.e., 
\begin{equation}
    \ten{F}^{(\beta)}
    =
    \ten{R}^{\top}
    \,\ten{F}\,
    \ten{R},
    \qquad
    \text{with }
    \ten{R}
    =
    \begin{bmatrix}
    \cos\beta & -\sin\beta & 0\\
    \sin\beta & \cos\beta  & 0\\
    0          & 0           & 1
    \end{bmatrix},
\end{equation}
where $\ten F^{(\beta)}$ represents the deformation gradient expressed in the rotated loading frame. We focus in particular on the representative case $\beta = 45^\circ$, which introduces a balanced coupling between the two in-plane directions and therefore departs most clearly from the axis-aligned loading states used during training. For this choice, the rotated deformation gradient takes the explicit form
\begin{equation}
    \ten{F}^{(45)}
    =
    \begin{bmatrix}
    \dfrac{\lambda_1+\lambda_2}{2} & \dfrac{\lambda_2-\lambda_1}{2} & 0\\[4pt]
    \dfrac{\lambda_2-\lambda_1}{2} & \dfrac{\lambda_1+\lambda_2}{2} & 0\\[4pt]
    0 & 0 & \left(\lambda_1\lambda_2\right)^{-1}
    \end{bmatrix},
    \label{eq:F_45deg}
\end{equation}
with the corresponding invariants computed as
\begin{equation}
    I_1^{(45)}
    =
    \lambda_1^2
    +
    \lambda_2^2
    +
    \left(\lambda_1 \lambda_2\right)^{-2},
    \qquad
    I_4^{(45)}
    =
    \frac{\lambda_1^2+\lambda_2^2}{2},
    \qquad
    I_6^{(45)}
    =
    \frac{\lambda_1^2+\lambda_2^2}{2}.
\end{equation}
Notably, $I_4^{(45)}$ and $I_6^{(45)}$ coincide, reflecting the symmetric contribution of the two preferred material directions under a $45^\circ$ rotation. This condition therefore probes the inferred anisotropic response encoded by the trained surrogate in a configuration where the directional invariants become indistinguishable, i.e., $I_4^{(45)}=I_6^{(45)}$. Consequently, the rotated loading does not independently activate the two anisotropic invariant contributions, but instead provides an off-axis consistency check of the coupled constitutive response under a loading orientation not explicitly included during training.

In addition, the rotated loading introduces nonzero off-diagonal terms in $\ten F^{(45)}$ and therefore induces in-plane shear. In general, the first Piola–Kirchhoff stress is not symmetric. In the present setting, however, the unit cells exhibit orthogonal symmetry with respect to the material axes, and the prescribed macroscopic deformation gradient is symmetric, corresponding to pure stretch without superposed rigid body rotation. For the loading cases considered here, we consistently observe that the homogenized nominal stress is symmetric, with $P_{12}=P_{21}$ up to machine precision. We therefore restrict attention to the independent in-plane components $P_{11}$, $P_{22}$, and $P_{12}$ as targets.

\subsection{Data-efficient surrogate training via active learning}
\label{sec:res_active}

Direct surrogate training in image space is computationally prohibitive for microstructures $\vct{M}_i\in\mathbb{R}^{9216}$. Accordingly, we apply the feature engineering framework described in \cref{sec:feature_engineering} to obtain a low-dimensional representation. For the present solid--void design set, the local-state indices entering the correlation functions take values $h,h'\in\{1,2\}$, where label $1$ denotes the solid phase and label $2$ denotes the solid--void interface extracted through periodic nearest-neighbor convolution \citep{danesh2025reduced}. Following our previous work on the same design set \citep{danesh2025reduced}, we retain the solid--solid and interface--interface autocorrelations, $c_{\vct{r}}^{11,(i)}$ and $c_{\vct{r}}^{22,(i)}$, defined by \cref{eq:feature_correlation}. Since the interface pixels are defined through nearest-neighbor convolution, their two-point correlations effectively incorporate a restricted subset of four-point spatial statistics, thereby enriching the representation beyond conventional two-point statistics \citep{montes2018reduced}. After variance balancing of the two autocorrelation fields, PCA is applied to obtain a reduced representation. Based on the sensitivity analysis in \ref{app:pc_dimension}, which shows that the predictive error reaches a plateau at six retained PC scores, we set $n_z=6$ for the present application. The resulting representation reduces the input dimensionality from $n_p=9,216$ to $n_z=6$ while preserving the essential morphological variability.

Recalling the relation $\vct{\theta}_i(\vct{\xi}_i(\vct{z}_i))$, we construct a probabilistic GP surrogate for the map $\vct{z}_i\in\mathbb{R}^6\mapsto \vct{\xi}_i\in\mathbb{R}^3$ and use the positivity transformation \cref{eq:pos_map} to compute effective parameters $\vct{\theta}_i\in\mathbb{R}^{3}$. To capture the correlation among these outputs, we employ a multi-output GP with LMC structure. We use $n_r=6$ latent GPs in total to flexibly capture correlations among the three effective parameters.

Prior to GP training, the reduced inputs are standardized to have zero mean and unit variance using the mean $\vct{\upmu}_z\in\mathbb{R}^{n_z}$ and standard deviation $\vct{\upsigma}_z\in\mathbb{R}^{n_z}$ computed over the full design set $\mathcal{D}$. The standardized input features are therefore given by
\begin{equation}
    \widehat{\vct{z}}_i
    =
    \left(\vct{z}_i-\vct{\upmu}_z\right)\oslash \vct{\upsigma}_z,
    \qquad \forall\,i \in \mathcal{I},
\end{equation}
where $\oslash$ denotes componentwise division. The stress observations are also standardized in the same manner, with separate statistics computed for each stress component at each sampled deformation state using the initial training set $\mathcal{T}_0$. Recalling the observation vector $\vct{y}^{(i)}_j\in\mathbb{R}^{n_{\mathrm{obs}}}$ introduced in \cref{eq:obs_model}, we denote by $\vct{\upmu}_{y,j}\in\mathbb{R}^{n_{\mathrm{obs}}}$ and $\vct{\upsigma}_{y,j}\in\mathbb{R}^{n_{\mathrm{obs}}}$ the mean and standard deviation computed from the initial labeled set $\mathcal{T}_0$ for each fixed deformation state $j\in\{1,\dots,n_f\}$. The normalized observations are then written as
\begin{equation}
    \widehat{\vct{y}}^{(i)}_j
    =
    \Big(
    \vct{y}^{(i)}_j - \vct{\upmu}_{y,j}
    \Big)
    \oslash
    \vct{\upsigma}_{y,j}.
\end{equation}
The normalization factors are then held fixed during active learning, and each newly acquired observation is standardized before being added to the training set. This keeps the observed responses on comparable numerical scales across the full loading history and supports stable GP training.

The model is implemented in GPyTorch and optimized using Adam with cosine-annealing learning rate scheduling and warm restarts at each active learning retraining step. The ELBO objective \cref{eq:elbo_loss} is approximated by Monte Carlo sampling for the expected log-likelihood term using $S=64$ samples in \cref{eq:mc_elbo}; unless stated otherwise, the same sample count is used for all other Monte Carlo estimates reported throughout this work. An initial labeled training set of size $|\mathcal{T}_0|=10$ and a fixed hold-out set of size $|\mathcal{H}|=500$ are generated via Latin hypercube sampling, with the latter reserved to monitor out-of-sample performance through the MAE defined in \cref{eq:al_mae}.

To train the surrogate with as little oracle-labeled data as possible, we adopt the active learning workflow described in \cref{sec:active_learning}. Starting from the initial set $\mathcal{T}_0$, the surrogate is first trained by maximizing the ELBO via \cref{eq:elbo_opt}. At each subsequent iteration, the predictive uncertainty over the unlabeled pool is quantified using the acquisition criterion in \cref{eq:al_acq}. The most informative candidate is selected according to \cref{eq:al_select}, labeled by the high-fidelity oracle via \cref{eq:al_oracle}, and appended to the training set following \cref{eq:al_update}. The surrogate is then retrained, and the procedure is repeated up to $T_{\max}=210$ iterations. Convergence is monitored using the stopping rule in \cref{eq:al_stop} with $\epsilon=10^{-3}$ and $L=5$.

\begin{figure}[h]
    \centering
    \includegraphics{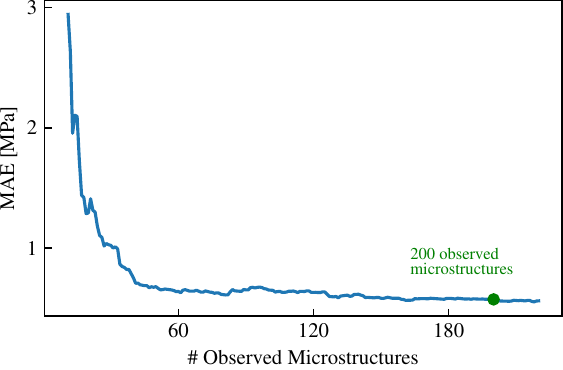}
    \caption{Active learning curve showing the MAE over the hold-out test set as a function of the number of observed microstructures. The error decreases rapidly and saturates after approximately 200 labeled designs.}
    \label{fig:al_curve}
\end{figure}

The active learning curve is shown in \cref{fig:al_curve}, where the hold-out MAE decreases rapidly during the initial iterations and saturates after approximately 200 observed microstructures. Although this indicates that the stopping criterion \cref{eq:al_stop} is already satisfied around 200 observations, we continue the iterations to $T_{\max}=210$ to show the saturation behavior more clearly. Relative to the full design space of size $n_d=50,000$, this small number of observations corresponds to 0.4\% labeled samples, demonstrating strong data efficiency. The surrogate trained with 200 observations is therefore adopted for the subsequent design selection study. For this surrogate, the jointly learned homoscedastic noise variance is $\sigma^2=2.83\times10^{-3}$ in the standardized stress space. This small magnitude is expected for high-fidelity data generated by established simulation protocols \citep{kalidindi2020feature}, such as the FFT-based homogenization adopted here.

\begin{figure}[h]
    \centering
    \includegraphics{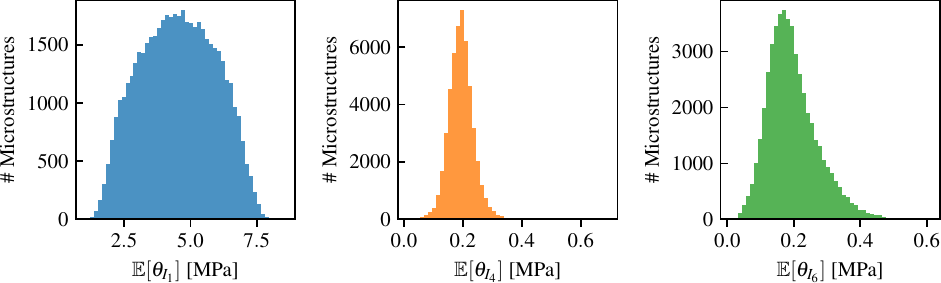}
    \caption{Empirical distributions of the Monte Carlo estimates of the posterior expected constitutive parameters 
    $\mathbb{E}[\theta_{I_1}]$, $\mathbb{E}[\theta_{I_4}]$, and $\mathbb{E}[\theta_{I_6}]$ 
    across the design set for the surrogate trained with 200 observations.}
    \label{fig:latent_param}
\end{figure}

The empirical distribution of the inferred parameters from the surrogate trained with 200 observations is shown in \cref{fig:latent_param}. Here, the plotted quantities correspond to Monte Carlo estimates of the posterior expectations of the effective parameters, obtained by sampling from the predictive distribution of the latent variables and propagating these samples through the positivity map \cref{eq:pos_map}. The distribution of $\mathbb{E}[\theta_{I_1}]$ is centered at larger values than those of $\mathbb{E}[\theta_{I_4}]$ and $\mathbb{E}[\theta_{I_6}]$, indicating that the isotropic contribution dominates the effective response. This observation is consistent with the Neo-Hookean solid phase behavior used in the high-fidelity oracle (cf.~\cref{sec:res_Oracle}). Although the directional parameters associated with $I_4$ and $I_6$ lie within comparable ranges, their empirical distributions differ in shape, indicating statistically distinct directional responses across the design set. Since the adopted biaxial loading family is symmetric with respect to the two in-plane directions, these differences cannot be attributed to directional bias in the loading protocol and instead reflect genuinely direction-dependent (orthotropic) contributions in the inferred effective response.

\subsection{Performance of the design selection framework}
\label{sec:res_inverse}
\begin{figure}[h]
    \centering
    \includegraphics{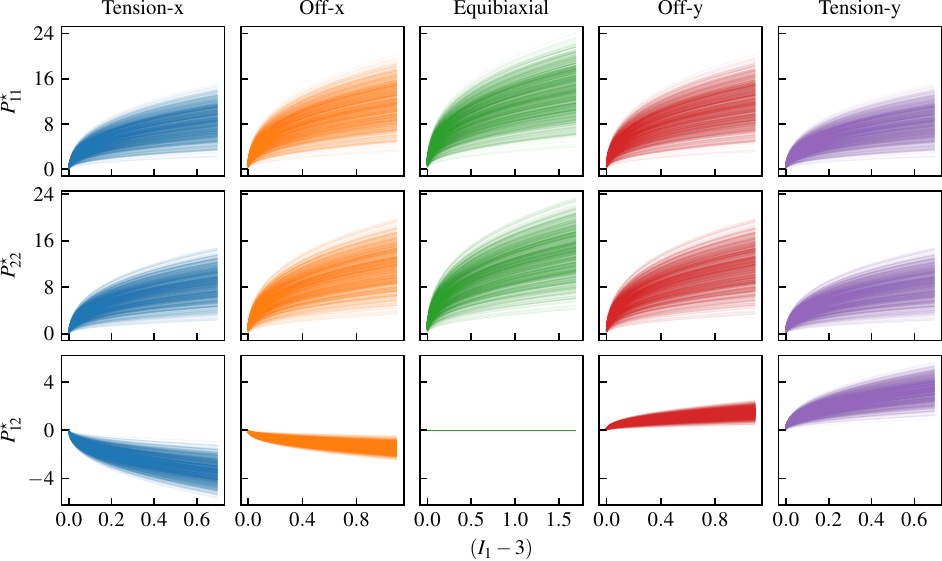}
    \caption{Stress responses of the $1,000$ design selection targets across loading paths. Columns correspond to loading paths, and rows display the in-plane stress components $P_{11}^{\star}$, $P_{22}^{\star}$, and $P_{12}^{\star}$ as functions of $(I_1-3)$.}
    \label{fig:target_stress}
\end{figure}

After training the surrogate, it can be employed to guide design selection toward promising regions of the discrete design space while minimizing the number of expensive forward oracle evaluations. To evaluate the performance of the proposed design selection framework, we select $N_{\mathrm{tar}} = 1,000$ structures from the discrete design set, ensuring that none of them were observed during surrogate training. These structures therefore serve as genuinely unseen targets. For each target, the forward oracle is evaluated under the rotated deformation gradient $\ten F^{(45)}$, yielding the three in-plane stress components $P_{11}^\star$, $P_{22}^\star$, and $P_{12}^\star$, whose trajectories along the prescribed loading paths are summarized in \cref{fig:target_stress}. Using these responses, we define objectives based on all seven possible combinations of the three stress components. Since each of the $1,000$ structures is considered with each target combination, this setup results in a total of $7,000$ design selection tasks, providing a statistically meaningful basis for performance assessment. The design selection study employs the uncertainty-aware score in \cref{eq:score}, with the automatically balanced weight $\lambda=\overline{\mu}/\overline{\sigma}$. The sensitivity of the selection performance to this choice of $\lambda$ is examined in \ref{app:lambda_sensitivity}. In the aggregated error \cref{eq:inv_nmae_bar}, we set $\omega_p=1$ for all target components, i.e., no stress component is prioritized over another. Unless stated otherwise, the error threshold and oracle evaluation budget are fixed at $\eta=5\%$ and $E_{\max}=50$, respectively, throughout this section.

\begin{table}
\centering
\caption{Design selection performance for all seven target stress component combinations, reporting threshold hit rates $R_{\eta}^{(\le E_{\mathrm{hit}})}$ and corresponding $R^2$ values.}
\label{tab:results_summary}
    \begin{tabular}{lccccc}
    \hline
    {Target} 
    & {$R_{\eta}^{(\le 50)}$ [\%]} 
    & {$R_{\eta}^{(\le 20)}$ [\%]} 
    & $R^2_{11}$ 
    & $R^2_{22}$ 
    & $R^2_{12}$ \\
    \hline
    $P^{\star}_{11}$ 
    & 100.0 & 99.9 & 0.992 & -- & -- \\

    $P^{\star}_{22}$ 
    & 100.0 & 99.8 & -- & 0.993 & -- \\

    $P^{\star}_{12}$ 
    & 99.7 & 98.7 & -- & -- & 0.987 \\

    $\{P^{\star}_{11},\,P^{\star}_{22}\}$ 
    & 98.8 & 97.8 & 0.989 & 0.990 & -- \\

    $\{P^{\star}_{11},\,P^{\star}_{12}\}$ 
    & 98.3 & 95.8 & 0.989 & -- & 0.982 \\

    $\{P^{\star}_{22},\,P^{\star}_{12}\}$ 
    & 99.1 & 96.5 & -- & 0.987 & 0.977 \\

    $\{P^{\star}_{11},\,P^{\star}_{22},\,P^{\star}_{12}\}$ 
    & 97.1 & 93.9 & 0.983 & 0.985 & 0.974 \\
    \hline
    \end{tabular}
\end{table}

\paragraph{\textbf{Budget-constrained threshold hit rates}} For each target, we record the oracle evaluation count $E_{\eta}$ required to reach the threshold $\eta$ from \cref{eq:inv_Eeta}. Using this quantity, we evaluate the threshold hit rate $R_{\eta}^{(\le E_{\mathrm{hit}})}$ at any reporting checkpoint $E_{\mathrm{hit}}\le E_{\max}$, 
\begin{equation}
    R_{\eta}^{(\le E_{\mathrm{hit}})}
    =
    \frac{1}{N_{\mathrm{tar}}}
    \sum_{k=1}^{N_{\mathrm{tar}}}
    \mathbb{I}\,\Big(E_{\eta}^{(k)} \le E_{\mathrm{hit}}\Big),
\end{equation}
where $E_{\eta}^{(k)}$ is the number of oracle evaluations required to reach $\eta$ for target $k$. Moreover, $\mathbb{I}$ denotes the indicator function, counting the number of targets that fulfill the condition $E_{\eta}^{(k)} \le E_{\mathrm{hit}}$. We report the threshold hit rate $R_{\eta}^{(\le E_{\mathrm{hit}})}$ at the two checkpoints $E_{\mathrm{hit}}=20$ and $E_{\mathrm{hit}}=50$ in \cref{tab:results_summary}, demonstrating that the design selection procedure remains highly effective across all seven target combinations. By the full allowed budget $E_{\max}=50$ (equivalently, at the reporting checkpoint $E_{\mathrm{hit}}=50$), the threshold hit rate exceeds 97.1\% in every case, while even after only 20 oracle evaluations it remains above 93.9\% for all combinations. As expected, the most restrictive objective, namely matching all three components simultaneously, is the most demanding and, therefore, exhibits the lowest hit rates within a limited oracle budget. Conversely, when the objective is restricted to a single stress component, the design selection problem becomes less constrained, which is reflected in higher threshold hit rates even at limited oracle budgets. For instance, for the single component targets $P_{11}^\star$ and $P_{22}^\star$ (considered separately), the hit rate is already almost 100\% at the earlier reporting checkpoint $E_{\mathrm{hit}}=20$. A clear trend is that objectives including the shear component $P_{12}^\star$ are generally more challenging than objectives based only on $P_{11}^\star$ and $P_{22}^\star$. This behavior is consistent with the training setup, where the surrogate was calibrated on axis-aligned loading responses $P_{11}$ and $P_{22}$, and did not directly observe the shear component $P_{12}$ during training. Taken together, the consistently high hit rates across all target combinations indicate that accurate design selection can be achieved with remarkably few oracle evaluations, underscoring the effectiveness of the Bayesian-guided design selection strategy.

\begin{figure}[h]
    \centering
    \includegraphics{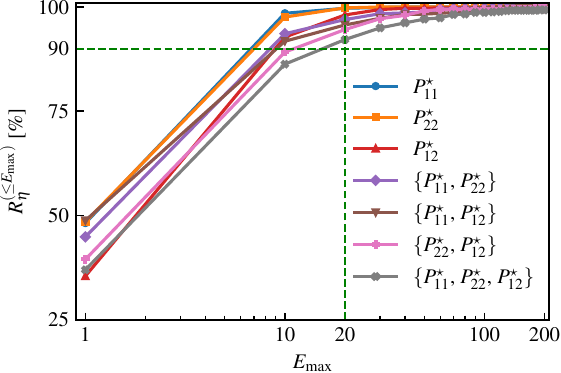}
    \caption{Threshold hit rate $R_{\eta}^{(\le E_{\max})}$ as a function of the oracle evaluation budget $E_{\max}$ for all seven target stress component combinations.}
    \label{fig:hit_rate}
\end{figure}

\paragraph{\textbf{Threshold hit rate vs. budget}} We examine how the design selection performance evolves as the allowed oracle evaluation budget is increased. The corresponding behavior is shown in \cref{fig:hit_rate}, where the threshold hit rate $R_{\eta}^{(\le E_{\max})}$ is plotted as a function of oracle budget $E_{\max}$ for all seven target combinations. For each case, the hit rate increases monotonically with the available budget and approaches 100\% in the large-budget regime. Already by $E_{\max}=200$, all curves lie very close to full hit rate. This trend indicates that the framework is not fundamentally limited by the absence of suitable designs within the discrete set. Rather, at smaller budgets, performance is primarily constrained by the number of oracle evaluations available to identify them. Therefore, by allowing for a sufficient number of oracle evaluations, the framework progressively reduces the gap between the identified and optimal designs within the discrete set. In the limit of sufficiently large budgets, the probability of finding a structure that satisfies the prescribed accuracy threshold approaches unity for all considered target combinations. This behavior confirms that the proposed Bayesian-guided design selection procedure scales predictably with the available computational budget and provides a controllable trade-off between evaluation cost and design accuracy.

\begin{figure}[h]
    \centering
\includegraphics{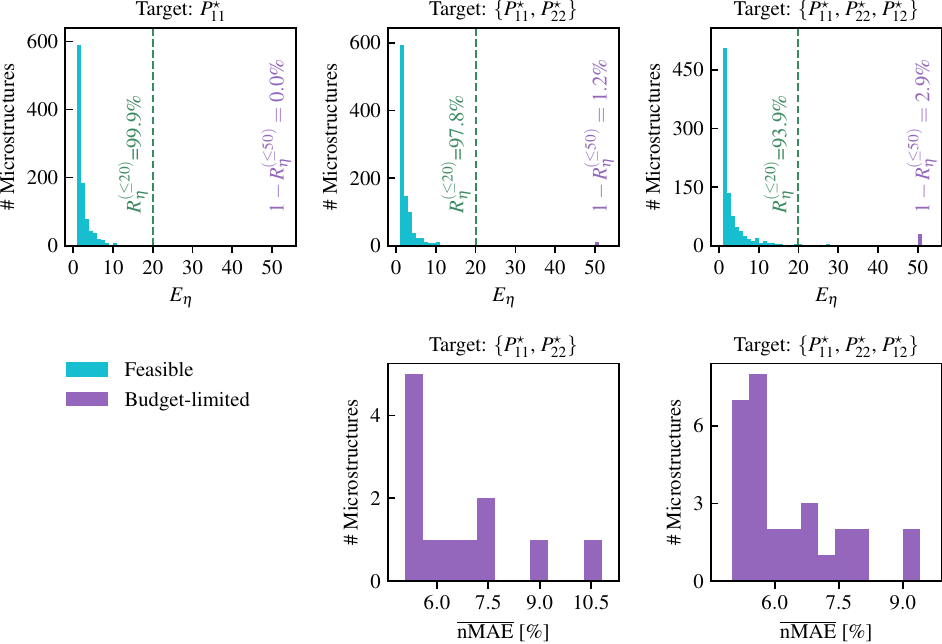}
    \caption{Distribution of oracle evaluations required to reach the prescribed threshold $\eta = 5\%$ for the selected target combinations $P_{11}^{\star}$, $\{P_{11}^{\star},P_{22}^{\star}\}$, and $\{P_{11}^{\star},P_{22}^{\star},P_{12}^{\star}\}$ (top row), and distribution of $\overline{\mathrm{nMAE}}$ for cases in which the oracle budget is exhausted without meeting the threshold (bottom row).}
    \label{fig:oracle_eval_hist}
\end{figure}

\paragraph{\textbf{Oracle evaluations to reach a prescribed threshold}} To further investigate the performance of the Bayesian-guided design selection procedure, we analyze the distribution of oracle evaluations required to reach the prescribed threshold. To avoid an excessive number of figures, the detailed diagnostics are restricted to three representative target sets: $P_{11}^\star$, $\{P_{11}^\star,P_{22}^\star\}$, and $\{P_{11}^\star,P_{22}^\star,P_{12}^\star\}$. The distribution of oracle evaluations in \cref{fig:oracle_eval_hist} shows that, for the majority of targets, the threshold is attained within fewer than 10 oracle evaluations, indicating rapid convergence of the design selection process in most cases. Only a small fraction of tasks does not reach the prescribed threshold within the available budget (i.e., $E_{\max}=50$), namely approximately 1.2\% for $\{P_{11}^\star,P_{22}^\star\}$ and 2.9\% for $\{P_{11}^\star,P_{22}^\star,P_{12}^\star\}$. Importantly, the lower panels of the same figure show that even these budget-limited optima, obtained from \cref{eq:inv_final}, remain close to their respective targets, as the aggregated error $\overline{\mathrm{nMAE}}$, defined in \cref{eq:inv_nmae_bar}, remains moderate and below approximately 11\%. Thus, even when the threshold is not reached within the allowed oracle budget, the returned inverse-designed solution remains close to the target, with only a moderate error.

\paragraph{\textbf{Computational efficiency}}
Computational efficiency arises from assigning the screening of the large library to the surrogate and reserving the oracle for final verification. Uncertainty-driven active learning limits the one-time high-fidelity training set to only 200 structures, or $0.4\%$ of the library of 50,000 structures. The resulting surrogate is reused for all 7,000 design tasks without retraining: mean predictions over the complete library are evaluated in batches, target mismatches are vectorized, and predictive uncertainty is evaluated only for the 50 shortlisted candidates. These inexpensive operations rapidly reduce the full library to a small ordered set for oracle evaluation. Moreover, $E_{\max}=50$ is a conservative safeguard rather than the typical demand. Across all 7,000 tasks, the online oracle count had a median of 1 (interquartile range 1--3) and a mean of 3.5; $93.8\%$ of the tasks reached $\eta$ within 10 evaluations, and $99.0\%$ did so within the full budget. The mean per-task cost therefore corresponds to evaluating just $0.007\%$ of the library.

\begin{figure}[!h]
    \centering
    \includegraphics{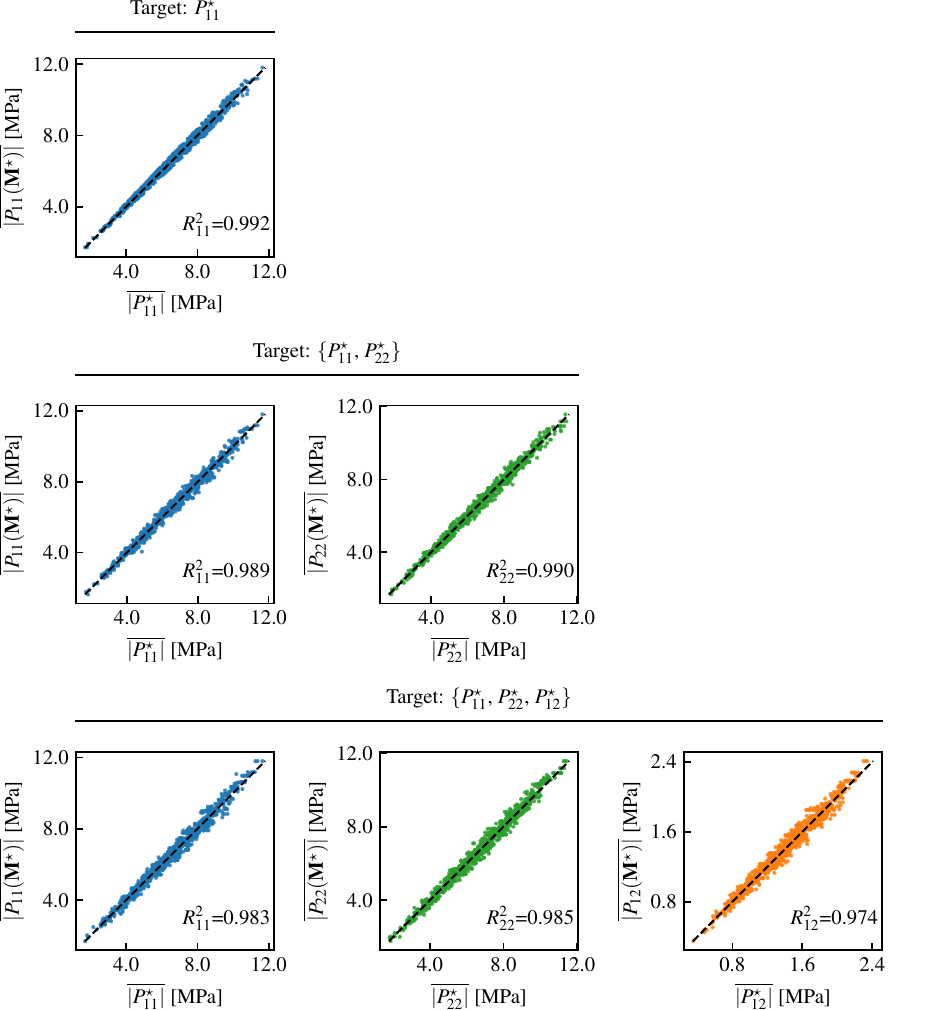}
    \caption{Parity plots comparing the mean absolute stresses of the inverse-designed structures with the corresponding target stresses for the selected component combinations $P_{11}^{\star}$, $\{P_{11}^{\star},P_{22}^{\star}\}$, and $\{P_{11}^{\star},P_{22}^{\star},P_{12}^{\star}\}$. The mean is computed over all loading paths, and both threshold-met and budget-limited solutions are included.}
    \label{fig:parity}
\end{figure}

\paragraph{\textbf{Achieved vs. target stress response}} 
To directly assess how closely the selected structures reproduce the prescribed stress responses, we compare the achieved stresses against their corresponding targets for $E_{\max}=50$. The accompanying coefficients of determination reported in \cref{tab:results_summary}, i.e., $R^2$ scores, quantify the agreement between the target and achieved responses. Specifically, $R^2_{11}$, $R^2_{22}$, and $R^2_{12}$ refer to the parity quality of the achieved vs. target responses for $P_{11}$, $P_{22}$, and $P_{12}$, respectively. For each active stress component, these values are computed from the mean absolute stress response over all loading paths. Consistent with the observations for the threshold hit rates, objectives including the shear component $P_{12}^\star$, which was not observed during training, generally exhibit slightly lower $R^2$ values than objectives involving only $P_{11}^\star$ and $P_{22}^\star$, reflecting the increased difficulty associated with matching the shear response. Nevertheless, the shear component $P_{12}$ still achieves good agreement with the targets and a reasonably high $R^2$ score. To illustrate the agreement between targets and achieved responses, parity plots of the mean absolute stress responses (over all loading paths), including both threshold-met and budget-limited solutions, are shown in \cref{fig:parity} for three representative target combinations $P_{11}^{\star}$, $\{P_{11}^{\star},P_{22}^{\star}\}$, and $\{P_{11}^{\star},P_{22}^{\star},P_{12}^{\star}\}$. The data cluster tightly around the diagonal, indicating strong agreement between the prescribed targets and the achieved responses, even for the previously unseen shear component $P_{12}$. Importantly, this consistency is maintained even when budget-limited solutions are included, confirming that the inverse-designed structures accurately reproduce the intended stress behavior across the considered objectives.

\subsection{Comparison with alternative design approaches}
\label{sec:res_baselines}

To place the design selection performance in context, we compare the proposed framework with random search and standard Bayesian optimization using expected improvement (BO--EI). All methods are evaluated on the same $N_{\mathrm{tar}}=1,000$ unseen targets for the most restrictive objective $\{P_{11}^{\star},P_{22}^{\star},P_{12}^{\star}\}$, over the complete pool of $50,000$ structures, with $\eta=5\%$ and at most $E_{\max}=50$ online oracle evaluations. The proposed framework retains its original surrogate trained on 200 active learning samples. We consider BO--EI using six PC scores obtained either from the proposed feature engineering strategy or directly from the flattened binary structure images; the latter variant provides a controlled assessment of the contribution of the proposed feature engineering. For a fair comparison, each BO–EI variant is initialized with a fixed set of 200 structures selected by Latin hypercube sampling in its respective feature space, ensuring that all model-based approaches have the same number of labeled structures before the online search. The initialization is disjoint from all targets and is reused for every target. As with the training labels of the proposed surrogate, these shared initialization labels are not counted toward the online per-target evaluation budget. For every target, BO--EI fits a scalar exact GP to the logarithm of the same design loss used by the proposed framework, as defined in \cref{eq:inv_loss}, with the same ARD-SE kernel \cref{eq:ard_se_kernel}; it updates the posterior after each online oracle evaluation and selects the next structure by maximizing expected improvement. Random search draws candidates uniformly without replacement, using one fixed realization per target.

\begin{figure}[!h]
    \centering
    \includegraphics[width=\linewidth]{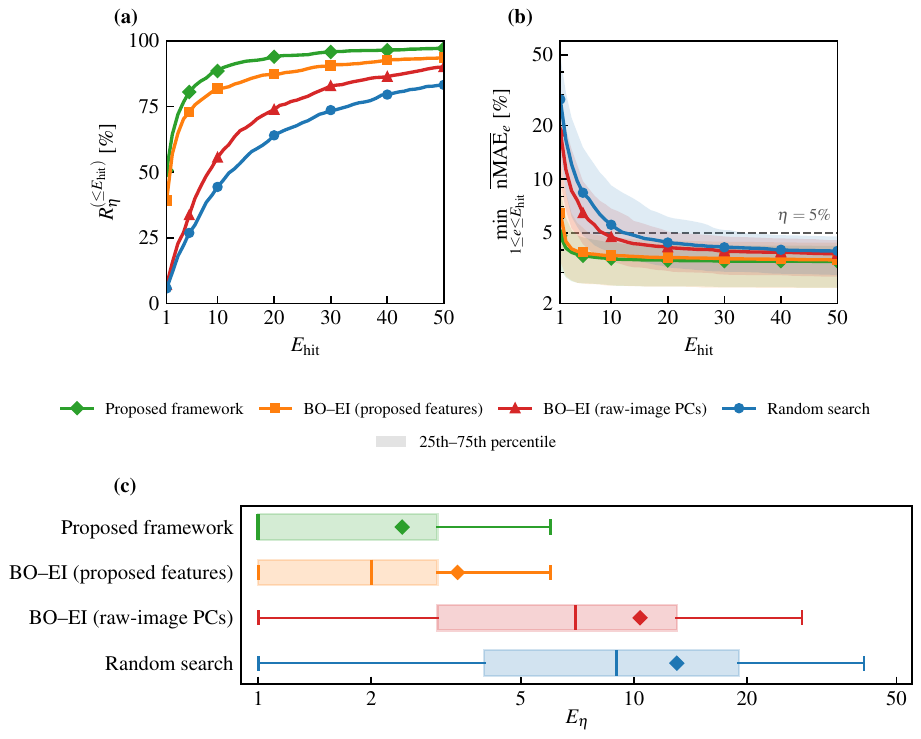}
    \caption{Comparison of the proposed framework, BO--EI with the proposed engineered features, BO--EI with raw-image PC features, and random search over the same 1,000 unseen targets. (a) Threshold hit rate as a function of the online oracle evaluation count. (b) Median best-so-far aggregated error; shaded regions denote the 25th--75th percentile range. (c) Distribution of the oracle evaluations required to reach the threshold for the 781 targets successfully resolved by all methods within $E_{\max}=50$. Boxes denote the interquartile range, center lines indicate medians, whiskers extend to the most extreme observations within 1.5 interquartile ranges, and diamonds indicate means.}
    \label{fig:inverse_baseline_comparison}
\end{figure}

\Cref{fig:inverse_baseline_comparison} compares the four methods from three complementary perspectives: panel (a) tracks the threshold hit rate with the online evaluation budget; panel (b) shows the median best-so-far error, with shading denoting the 25th--75th percentile range; and panel (c) uses box plots to show the distribution of $E_{\eta}$ for the 781 targets successfully resolved by all four methods within $E_{\max}=50$. The clearest separation occurs in the low-budget regime. For instance, at $E_{\mathrm{hit}}=1$, the proposed framework resolves $50.7\%$ of the targets, compared with $39.3\%$ for BO--EI using the engineered PC features, $7.1\%$ for BO--EI using raw-image PCs, and $5.8\%$ for random search. The two BO--EI variants follow the same optimization protocol, so their comparison assesses the effect of the structural representation under a common search procedure. More precisely, the PC scores derived from statistical correlations increase the hit rate more than fivefold with only one evaluation, demonstrating that they provide a more task-relevant representation for response-guided search in the present design set than direct PCA of the raw images. With the same features, the proposed framework still outperforms BO--EI. This advantage persists across the full evaluation budget, for which the proposed framework maintains the highest hit rate and lowest median best-so-far error. For these 781 commonly resolved targets, the median $E_{\eta}$ values are 1, 2, 7, and 9 for the proposed framework, BO--EI with the proposed features, BO--EI with raw-image PCs, and random search, respectively. Taken together, these results show that the performance gain arises jointly from the task-relevant feature representation and the complete surrogate-guided design selection framework.

\paragraph{\textbf{Remark}}
Topology optimization is a complementary approach that can synthesize a new unit cell geometry rather than select from a prescribed library. Closely related finite-strain studies demonstrate accurate matching of nonlinear mechanical responses, but require a separate target-specific optimization. \citet{thillaithevan2024inverse} reported errors below 3\% for targeted bidirectional softening, using prescribed budgets of 650–1050 design iterations across their examples. \citet{zhang2019computational} used on the order of 500 design iterations for finite-strain auxetic designs, while \citet{aveline2026topology} reported 92--331 design iterations for bistable examples, depending on the initialization. In contrast, after the common one-time offline training stage, the present surrogate is reused without retraining across all 7,000 design tasks. As quantified in \cref{sec:res_inverse}, the median and mean numbers of online oracle evaluations used are 1 and 3.5 per task, respectively, under $E_{\max}=50$. Although the work associated with an individual iteration or evaluation depends on the formulation and solver, these reported counts provide a practical measure of the target-specific search effort. The comparison therefore demonstrates the efficiency of the proposed framework for repeated design queries over an existing admissible library, while topology optimization retains the advantage of generating designs beyond that library.

\section{Conclusions}

In this work, we presented a data-efficient framework for screening and selection from large design sets of hyperelastic microstructures, combining probabilistic surrogate modeling, uncertainty-driven active learning, and Bayesian-guided exploration of a discrete design space. The proposed approach enables design selection within large pools of feasible stochastic structures while requiring only a minimal number of labeled samples and very few expensive high-fidelity simulations. In essence, a cheaply trained surrogate model identifies promising regions of the discrete design space, after which a small number of high-fidelity oracle evaluations are used to iteratively determine the structure that best matches a prescribed effective mechanical response.

More precisely, the framework targets structural design problems in which the geometry cannot be conveniently parameterized, and continuous optimization therefore becomes impractical. Instead of searching a continuous parameter space, the method operates on a discrete pool of feasible candidate structures while avoiding the need to evaluate all candidates using expensive simulations or experiments. A probabilistic surrogate is constructed to capture the structure–property relation using only a small number of labeled designs. To minimize the required training data, uncertainty-driven active learning is employed, in which the predictive uncertainty of the surrogate guides the selection of new structures to be evaluated and added to the training set. To ensure physical consistency and interpretability, the surrogate does not directly predict the effective stress response. Instead, it maps structural descriptors to the parameters of an effective constitutive model describing the macroscopic behavior of the hyperelastic microstructure. As structural representations are inherently high-dimensional, statistical feature engineering is used to extract a compact set of reduced descriptors that retain the essential structural information while keeping surrogate learning tractable. Once trained, the surrogate efficiently explores the candidate pool during the design selection stage and identifies promising structures. Predictive uncertainty is incorporated into the ranking of shortlisted candidates in order to penalize unreliable predictions. The final decision is then made through a small number of high-fidelity oracle evaluations, which ultimately determine the design that best matches the prescribed target response. Baseline comparisons showed that the proposed framework outperformed random search and BO--EI using either the same engineered features or raw-image PCs throughout the evaluated oracle budget, consistently achieving higher threshold hit rates and lower median errors. Importantly, since the surrogate serves only as a guide for navigating the design space, the overall framework remains robust even when the surrogate is imperfect, for instance, due to limited training data or an insufficiently expressive constitutive ansatz. In such cases, inaccuracies in the surrogate merely affect the efficiency of the search, since the final selection of the optimal design is always resolved by the high-fidelity oracle.

To investigate the performance of the proposed framework, we employed it for screening and selection from a design set of stochastic hyperelastic metamaterials comprising $50,000$ admissible candidate structures. In this setting, uncertainty-driven active learning proved particularly effective, making it possible to train the surrogate using only 0.4\% of the entire candidates while still achieving strong predictive performance. During the design selection stage, the framework consistently identified candidates that matched the prescribed targets within the predefined error threshold using only a small number of high-fidelity oracle evaluations. In most cases, the target was reached in fewer than 10 oracle calls, while the threshold hit rate already exceeded 93.9\% within 20 evaluations and surpassed 97.1\% within 50 evaluations across the considered target combinations. Moreover, as the oracle budget increased, the hit rate approached 100\%, demonstrating that the framework provides a clear and controllable trade-off between oracle cost and design accuracy. Importantly, the method also remains robust when the prescribed threshold cannot be met within the available budget. In such cases, it returns a budget-limited optimum, which was shown to remain close to the target and to incur only moderate error.

Although the present application considers two-dimensional plane-stress structures with known orthotropic symmetry, these choices are specific to the selected test case and do not restrict the general framework. Extension to three-dimensional voxelized structures is straightforward within the same feature engineering and surrogate learning workflow. Compressive and shear behavior can be incorporated by extending the training data with loading paths that sufficiently excite the associated constitutive parameters. When the symmetry class is not known a priori, it may be identified from geometric invariances or from the symmetry properties of the instantaneous constitutive tangent \citep{olive2022characterization}; the corresponding structural tensors and a minimal invariant basis can then be constructed systematically \citep{riemer2026construction}. These extensions change the loading protocol and constitutive parameterization, but not the active learning or design selection procedures.

The present framework also has several limitations that merit discussion. First, the design selection problem is posed over a known finite pool of candidate structures. Consequently, the framework can only identify the best design within this predefined set, and its success therefore depends on whether the candidate pool already spans sufficiently rich and relevant regions of the admissible design space. Second, the effective constitutive description relies on a prescribed hyperelastic ansatz. While this choice provides interpretability and enables efficient surrogate learning, it nevertheless introduces a modeling assumption regarding the admissible functional form of the constitutive response and may therefore limit expressiveness when the true material behavior requires a richer representation.

These limitations naturally point toward several promising extensions of the present framework. Rather than relying on a fixed library of candidate structures, future work could incorporate generative modeling approaches to construct larger and more diverse pools of feasible designs. In such a hybrid strategy, generative models, such as variational autoencoders \citep{kim2021exploration,nguyen2026deep}, generative adversarial networks \citep{hsu2021microstructure,nguyen2022synthesizing}, diffusion models \citep{lyu2024microstructure,baishnab20253d}, or related geometry generation techniques, would primarily serve as candidate generators that expand the accessible design set, while the design selection task itself would remain a discrete surrogate-guided search validated by high-fidelity oracle evaluations. In parallel, the constitutive modeling component could evolve beyond a fixed hyperelastic ansatz toward model discovery frameworks, in which suitable constitutive forms are automatically identified from a richer library of admissible terms \citep{flaschel2021euclid,linka2023new,urreaquintero2026modeldiscovery,narouie2026unsupervised,anton2025uncertainty}. Such approaches could allow constitutive relations to be discovered either for individual structures or for families of structures grouped according to their symmetry class, thereby yielding more expressive and better adapted surrogate representations. More broadly, the overall framework is not restricted to hyperelastic materials and could be extended to inelastic settings \citep{ha2023rapid,zeng2023inverse,danesh2025two}, where history-dependent constitutive behavior must also be learned and incorporated within the design selection process. Beyond solid mechanics, the proposed methodology provides a general blueprint for data-efficient candidate screening and design selection across a wide range of scientific and engineering domains, provided that the underlying problem can be formulated within a similar surrogate--oracle framework.

\appendix
\renewcommand{\thefigure}{\Alph{section}.\arabic{figure}}
\renewcommand{\theHfigure}{appendix.\Alph{section}.\arabic{figure}}
\setcounter{figure}{0}

\section{Sensitivity to the retained PCA dimension}
\label{app:pc_dimension}

To identify an adequate number of retained PC scores for the present hyperelastic response, we conduct a dimensionality study. A fixed random subset of 5,000 labeled structures is partitioned once using an 80/20 train/test split, which is kept unchanged for all retained dimensions. To keep GP training scalable for this larger labeled set, all models are trained as sparse variational GPs (SVGPs) using 200 inducing inputs defined by the same fixed subset of training structures \citep{titsias2009variational,hensman2013gaussian}. We train the multi-output SVGP with between one and eight retained PC scores and evaluate the stress MAE on the common hold-out test set.

\begin{figure}[H]
    \centering
    \includegraphics{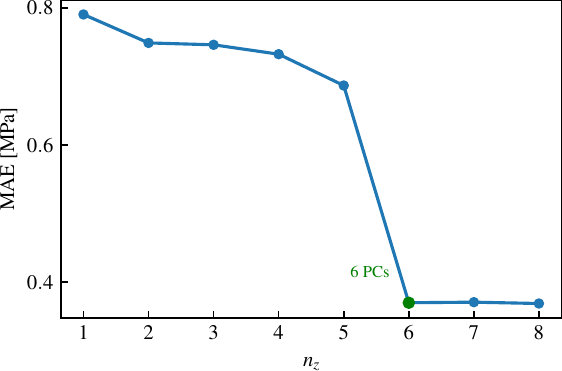}
    \caption{MAE evaluated on the fixed hold-out test set as a function of the number of retained PC scores. All models use the same fixed 80/20 train/test split. The error reaches a plateau at six PC scores.}
    \label{fig:pc_dimension_curve}
\end{figure}

As shown in \cref{fig:pc_dimension_curve}, the test MAE decreases sharply up to six PC scores and then plateaus, with no meaningful improvement thereafter. We therefore select $n_z=6$ as the smallest retained dimension at the onset of this plateau, avoiding unnecessary inputs without a meaningful loss of predictive accuracy.

\section{Sensitivity to the uncertainty weight}
\label{app:lambda_sensitivity}
\setcounter{figure}{0}

We assess the influence of the uncertainty weight $\lambda$ in \cref{eq:score} by writing $\lambda=\gamma\,\overline{\mu}/\overline{\sigma}$ and considering $\gamma\in\{0,0.25,0.5,1,2,4\}$. The case $\gamma=0$ ranks the fixed shortlist using only the predictive mean, while $\gamma=1$ corresponds to the automatically balanced value used in the proposed framework. The same 1,000 targets, shortlist size $E_{\max}=50$, Monte Carlo estimates, and oracle responses are used for every value of $\gamma$, so that only the ranking within the common shortlist changes.

\begin{figure}[H]
    \centering
    \includegraphics{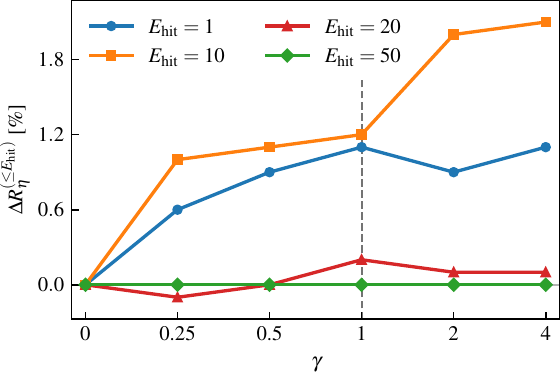}
    \caption{Improvement in threshold hit rate $\Delta R_{\eta}^{(\le E_{\mathrm{hit}})}$ relative to mean-only ranking (i.e., $\gamma=0$) as a function of the uncertainty weight multiplier $\gamma$. Curves report the performance at $E_{\mathrm{hit}}\in\{1,10,20,50\}$, and the vertical line at $\gamma=1$ marks the automatically balanced value used in the proposed framework.}
    \label{fig:lambda_sensitivity}
\end{figure}

At each reporting checkpoint $E_{\mathrm{hit}}$, the improvement over the mean-only ranking (i.e., $\gamma=0$) is defined as
\begin{equation}
    \Delta R_{\eta}^{(\le E_{\mathrm{hit}})}(\gamma)
    =
    R_{\eta}^{(\le E_{\mathrm{hit}})}(\gamma)
    -
    R_{\eta}^{(\le E_{\mathrm{hit}})}(0).
    \label{eq:lambda_hit_rate_improvement}
\end{equation}
Here, $R_{\eta}^{(\le E_{\mathrm{hit}})}(\gamma)$ is the threshold hit rate for the uncertatinty weight multiplier $\gamma$, so that $\Delta R_{\eta}^{(\le E_{\mathrm{hit}})}>0$ indicates an improvement over mean-only ranking. \cref{fig:lambda_sensitivity} shows that uncertainty weighting improves the hit rate in the low-budget regime. The automatically balanced choice (i.e., $\gamma=1$) provides this benefit without additional tuning, supporting its adoption in the framework. At $E_{\mathrm{hit}}=50$, the hit rate is insensitive to $\gamma$ since with $E_{\max}=50$, every structure in the shortlist has been evaluated and the ordering no longer matters.

\section*{CRediT authorship contribution statement}
\textbf{Hooman Danesh:} Conceptualization, Methodology, Software, Validation, Formal analysis, Investigation, Data Curation, Writing - Original Draft, Writing - Review \& Editing, Visualization
\textbf{Henning Wessels:} Conceptualization, Methodology, Resources, Writing - Review \& Editing, Supervision, Funding acquisition

\section*{Declaration of competing interest}
The authors declare that they have no known competing financial interests or personal relationships that could have appeared to influence the work reported in this paper.

\section*{Acknowledgments}
The authors gratefully acknowledge the financial support of zukunft.niedersachsen. Hooman Danesh would like to thank Jorge-Humberto Urrea-Quintero for helpful discussions and valuable suggestions.

\section*{Data availability}
The data and source code supporting the findings of this study is publicly available on Zenodo \citep{danesh2026BayGDS} and on GitHub at \url{https://github.com/hooman-danesh/BayGDS}.

\section*{Declaration of generative AI and AI-assisted technologies}
During the preparation of this work, the authors used ChatGPT (OpenAI) to improve the readability and clarity of the language. After using this tool, the authors reviewed and edited the content as needed and take full responsibility for the content of the published article.







\bibliographystyle{elsarticle-num-names}
\bibliography{references.bib}

@article{danesh2024fft,
  title={{FFT}-based surrogate modeling of auxetic metamaterials with real-time prediction of effective elastic properties and swift inverse design},
  author={Danesh, Hooman and Di Lorenzo, Daniele and Chinesta, Francisco and Reese, Stefanie and Brepols, Tim},
  journal={Materials \& Design},
  volume={248},
  pages={113491},
  year={2024},
  doi={10.1016/j.matdes.2024.113491},
  publisher={Elsevier}
}

@article{bastek2023inverse,
  title={Inverse design of nonlinear mechanical metamaterials via video denoising diffusion models},
  author={Bastek, Jan-Hendrik and Kochmann, Dennis M},
  journal={Nature Machine Intelligence},
  volume={5},
  number={12},
  pages={1466--1475},
  year={2023},
  doi={10.1038/s42256-023-00762-x},
  publisher={Nature Publishing Group UK London}
}

@article{kuok2025bayesian,
  title={Bayesian generative kernel {Gaussian} process regression},
  author={Kuok, Sin-Chi and Yao, Shuang-Ao and Yuen, Ka-Veng and Yan, Wang-Ji and Girolami, Mark},
  journal={Mechanical Systems and Signal Processing},
  volume={227},
  pages={112395},
  year={2025},
  doi={10.1016/j.ymssp.2025.112395},
  publisher={Elsevier}
}

@inproceedings{chen2023identifiability,
  title={On the identifiability and interpretability of {Gaussian} process models},
  author={Chen, Jiawen and Mu, Wancen and Li, Yun and Li, Didong},
  booktitle={Advances in Neural Information Processing Systems},
  volume={36},
  pages={70267--70278},
  year={2023},
  doi={10.52202/075280-3079}
}

@article{abdessalem2017automatic,
  title={Automatic kernel selection for {Gaussian} processes regression with approximate {Bayesian} computation and sequential {Monte Carlo}},
  author={Ben Abdessalem, Anis and Dervilis, Nikolaos and Wagg, David J. and Worden, Keith},
  journal={Frontiers in Built Environment},
  volume={3},
  pages={52},
  year={2017},
  doi={10.3389/fbuil.2017.00052},
  publisher={Frontiers}
}

@article{kalidindi2011microstructure,
  title={Microstructure informatics using higher-order statistics and efficient data-mining protocols},
  author={Kalidindi, Surya R and Niezgoda, Stephen R and Salem, Ayman A},
  journal={Jom},
  volume={63},
  number={4},
  pages={34--41},
  year={2011},
  doi={10.1007/s11837-011-0057-7},
  publisher={Springer}
}

@article{kalidindi2020feature,
  title={Feature engineering of material structure for {AI}-based materials knowledge systems},
  author={Kalidindi, Surya R},
  journal={Journal of Applied Physics},
  volume={128},
  number={4},
  pages={041103},
  year={2020},
  doi={10.1063/5.0011258},
  publisher={AIP Publishing}
}

@article{brough2017materials,
  title={Materials knowledge systems in python—a data science framework for accelerated development of hierarchical materials},
  author={Brough, David B and Wheeler, Daniel and Kalidindi, Surya R},
  journal={Integrating materials and manufacturing innovation},
  volume={6},
  number={1},
  pages={36--53},
  year={2017},
  doi={10.1007/s40192-017-0089-0},
  publisher={Springer}
}

@article{niezgoda2011understanding,
  title={Understanding and visualizing microstructure and microstructure variance as a stochastic process},
  author={Niezgoda, Stephen R and Yabansu, Yuksel C and Kalidindi, Surya R},
  journal={Acta Materialia},
  volume={59},
  number={16},
  pages={6387--6400},
  year={2011},
  doi={10.1016/j.actamat.2011.06.051},
  publisher={Elsevier}
}

@article{niezgoda2013novel,
  title={Novel microstructure quantification framework for databasing, visualization, and analysis of microstructure data},
  author={Niezgoda, Stephen R and Kanjarla, Anand K and Kalidindi, Surya R},
  journal={Integrating Materials and Manufacturing Innovation},
  volume={2},
  number={1},
  pages={54--80},
  year={2013},
  doi={10.1186/2193-9772-2-3},
  publisher={Springer}
}

@techreport{settles2009active,
  title={Active Learning Literature Survey},
  author={Settles, Burr},
  institution={University of Wisconsin-Madison Department of Computer Sciences},
  number={TR1648},
  year={2009}
}

@article{liu2024active,
  title={Active learning for regression of structure--property mapping: the importance of sampling and representation},
  author={Liu, Hao and Yucel, Berkay and Ganapathysubramanian, Baskar and Kalidindi, Surya R and Wheeler, Daniel and Wodo, Olga},
  journal={Digital Discovery},
  volume={3},
  number={10},
  pages={1997--2009},
  year={2024},
  doi={10.1039/d4dd00073k},
  publisher={Royal Society of Chemistry}
}

@article{buzzy2025active,
  title={Active learning for the design of polycrystalline textures using conditional normalizing flows},
  author={Buzzy, Michael O and de Oca Zapiain, David Montes and Generale, Adam P and Kalidindi, Surya R and Lim, Hojun},
  journal={Acta Materialia},
  volume={284},
  pages={120537},
  year={2025},
  doi={10.1016/j.actamat.2024.120537},
  publisher={Elsevier}
}

@article{ozbayram2025batch,
  title={Batch active learning for microstructure--property relations in energetic materials},
  author={Ozbayram, Ozge and Olsen, Daniel and Annamaraju, Maruthi and Robertson, Andreas E and Venkatraman, Aditya and Kalidindi, Surya R and Zhou, Min and Graham-Brady, Lori},
  journal={Mechanics of Materials},
  volume={205},
  pages={105308},
  year={2025},
  doi={10.1016/j.mechmat.2025.105308},
  publisher={Elsevier}
}

@article{de2017finite,
  title={Finite strain {FFT}-based non-linear solvers made simple},
  author={De Geus, TWJ and Vond{\v{r}}ejc, J and Zeman, J and Peerlings, RHJ},
  journal={Computer Methods in Applied Mechanics and Engineering},
  volume={318},
  pages={412--430},
  year={2017},
  doi={10.1016/j.cma.2016.12.032},
  publisher={Elsevier}
}

@article{minh2020surrogate,
  title={A surrogate model for computational homogenization of elastostatics at finite strain using high-dimensional model representation-based neural network},
  author={Minh Nguyen-Thanh, Vien and Trong Khiem Nguyen, Lu and Rabczuk, Timon and Zhuang, Xiaoying},
  journal={International Journal for Numerical Methods in Engineering},
  volume={121},
  number={21},
  pages={4811--4842},
  year={2020},
  doi={10.1002/nme.6493},
  publisher={Wiley Online Library}
}

@article{lucarini2022adaptation,
  title={Adaptation and validation of {FFT} methods for homogenization of lattice based materials},
  author={Lucarini, Sergio and Cobian, Lucia and Voitus, A and Segurado, J},
  journal={Computer Methods in Applied Mechanics and Engineering},
  volume={388},
  pages={114223},
  year={2022},
  doi={10.1016/j.cma.2021.114223},
  publisher={Elsevier}
}

@article{danesh2023challenges,
  title={Challenges in two-scale computational homogenization of mechanical metamaterials},
  author={Danesh, Hooman and Brepols, Tim and Reese, Stefanie},
  journal={PAMM},
  volume={23},
  number={1},
  pages={e202200139},
  year={2023},
  doi={10.1002/pamm.202200139},
  publisher={Wiley Online Library}
}

@article{vondvrejc2014fft,
  title={An {FFT}-based Galerkin method for homogenization of periodic media},
  author={Vond{\v{r}}ejc, Jaroslav and Zeman, Jan and Marek, Ivo},
  journal={Computers \& Mathematics with Applications},
  volume={68},
  number={3},
  pages={156--173},
  year={2014},
  doi={10.1016/j.camwa.2014.05.014},
  publisher={Elsevier}
}

@article{vondvrejc2015guaranteed,
  title={Guaranteed upper--lower bounds on homogenized properties by {FFT}-based Galerkin method},
  author={Vond{\v{r}}ejc, Jaroslav and Zeman, Jan and Marek, Ivo},
  journal={Computer Methods in Applied Mechanics and Engineering},
  volume={297},
  pages={258--291},
  year={2015},
  doi={10.1016/j.cma.2015.09.003},
  publisher={Elsevier}
}

@article{willot2015fourier,
  title={Fourier-based schemes for computing the mechanical response of composites with accurate local fields},
  author={Willot, Fran{\c{c}}ois},
  journal={Comptes Rendus M{\'e}canique},
  volume={343},
  number={3},
  pages={232--245},
  year={2015},
  doi={10.1016/j.crme.2014.12.005},
  publisher={Elsevier}
}

@article{danesh2025two,
  title={A two-scale computational homogenization approach for elastoplastic truss-based lattice structures},
  author={Danesh, Hooman and Heu{\ss}en, Lisamarie and Mont{\'a}ns, Francisco J and Reese, Stefanie and Brepols, Tim},
  journal={Results in Engineering},
  volume={25},
  pages={103976},
  year={2025},
  doi={10.1016/j.rineng.2025.103976},
  publisher={Elsevier}
}

@article{generale2024inverse,
  title={Inverse stochastic microstructure design},
  author={Generale, Adam P and Robertson, Andreas E and Kelly, Conlain and Kalidindi, Surya R},
  journal={Acta Materialia},
  volume={271},
  pages={119877},
  year={2024},
  doi={10.1016/j.actamat.2024.119877},
  publisher={Elsevier}
}

@article{flaschel2021euclid,
  title={Unsupervised discovery of interpretable hyperelastic constitutive laws},
  author={Flaschel, Moritz and Kumar, Siddhant and De Lorenzis, Laura},
  journal={Computer Methods in Applied Mechanics and Engineering},
  volume={381},
  pages={113852},
  year={2021},
  doi={10.1016/j.cma.2021.113852}
}

@article{urreaquintero2026modeldiscovery,
  title={Automated constitutive model discovery by pairing sparse regression algorithms with model selection criteria},
  author={Urrea-Quintero, Jorge-Humberto and Anton, David and De Lorenzis, Laura and Wessels, Henning},
  journal={Computer Methods in Applied Mechanics and Engineering},
  volume={449},
  pages={118551},
  year={2026},
  doi={10.1016/j.cma.2025.118551}
}

@misc{bastek2023data,
  title={Inverse-design of nonlinear mechanical metamaterials via video denoising diffusion models: dataset and model checkpoints},
  author={Bastek, Jan-Hendrik},
  year={2023},
  publisher={ETHZ Research Collection},
  doi={10.3929/ethz-b-000629716}
}

@misc{danesh2025data,
  title={Reduced-order structure-property linkages for stochastic metamaterials (dataset)},
  author={Danesh, Hooman and Annamaraju, Maruthi and Brepols, Tim and Reese, Stefanie and Kalidindi, Surya},
  year={2025},
  publisher={Zenodo},
  doi={10.5281/zenodo.15302946}
}

@article{danesh2025reduced,
  title={Reduced-order structure-property linkages for stochastic metamaterials},
  author={Danesh, Hooman and Annamaraju, Maruthi and Brepols, Tim and Reese, Stefanie and Kalidindi, Surya R},
  journal={Physical Review Materials},
  volume={9},
  number={7},
  pages={075201},
  year={2025},
  doi={10.1103/8zzt-4b7z},
  publisher={APS}
}

@article{narouie2026unsupervised,
  title={Unsupervised Constitutive Model Discovery from Sparse and Noisy Data},
  author={Narouie, Vahab Knauf and Urrea-Quintero, Jorge-Humberto and Cirak, Fehmi and Wessels, Henning},
  journal={Computer Methods in Applied Mechanics and Engineering},
  volume={452},
  pages={118722},
  year={2026},
  doi={10.1016/j.cma.2025.118722},
  publisher={Elsevier}
}

@article{anton2025uncertainty,
  title={Uncertainty quantification in model discovery by distilling interpretable material constitutive models from Gaussian process posteriors},
  author={Anton, David and Wessels, Henning and R{\"o}mer, Ulrich and Henkes, Alexander and Urrea-Quintero, Jorge-Humberto},
  journal={arXiv preprint arXiv:2510.22345},
  year={2025},
  url={https://arxiv.org/abs/2510.22345}
}

@article{linka2023new,
  title={A new family of constitutive artificial neural networks towards automated model discovery},
  author={Linka, Kevin and Kuhl, Ellen},
  journal={Computer Methods in Applied Mechanics and Engineering},
  volume={403},
  pages={115731},
  year={2023},
  doi={10.1016/j.cma.2022.115731},
  publisher={Elsevier}
}

@article{mcculloch2024automated,
  title={Automated model discovery for textile structures: The unique mechanical signature of warp knitted fabrics},
  author={McCulloch, Jeremy A and Kuhl, Ellen},
  journal={Acta Biomaterialia},
  volume={189},
  pages={461--477},
  year={2024},
  doi={10.1016/j.actbio.2024.09.051},
  publisher={Elsevier}
}

@article{rassloff2025inverse,
  title={Inverse design of spinodoid structures using Bayesian optimization},
  author={Ra{\ss}loff, Alexander and Seibert, Paul and Kalina, Karl A and K{\"a}stner, Markus},
  journal={Computational Mechanics},
  volume={77},
  number={1},
  pages={275--296},
  year={2026},
  doi={10.1007/s00466-024-02587-w},
  publisher={Springer}
}

@article{otto2025data,
  title={Data-Driven Inverse Design of Spinodoid Architected Materials},
  author={Otto, Alexandra and Rosenkranz, Max and Kalina, Karl A and K{\"a}stner, Markus},
  journal={GAMM-Mitteilungen},
  volume={48},
  number={4},
  pages={e70008},
  year={2025},
  doi={10.1002/gamm.70008},
  publisher={Wiley Online Library}
}

@article{holzapfel2000new,
  title={A new constitutive framework for arterial wall mechanics and a comparative study of material models},
  author={Holzapfel, Gerhard A and Gasser, Thomas C and Ogden, Ray W},
  journal={Journal of elasticity and the physical science of solids},
  volume={61},
  number={1},
  pages={1--48},
  year={2000},
  doi={10.1023/A:1010835316564},
  publisher={Springer}
}

@book{holzapfel2002nonlinear,
  title={Nonlinear Solid Mechanics: A Continuum Approach for Engineering},
  author={Holzapfel, Gerhard A.},
  year={2000},
  publisher={Wiley},
  isbn={9780471823193}
}

@article{kim2021exploration,
  title={Exploration of optimal microstructure and mechanical properties in continuous microstructure space using a variational autoencoder},
  author={Kim, Yongju and Park, Hyung Keun and Jung, Jaimyun and Asghari-Rad, Peyman and Lee, Seungchul and Kim, Jin You and Jung, Hwan Gyo and Kim, Hyoung Seop},
  journal={Materials \& Design},
  volume={202},
  pages={109544},
  year={2021},
  doi={10.1016/j.matdes.2021.109544},
  publisher={Elsevier}
}

@article{nguyen2026deep,
  title={Deep learning-aided inverse design of porous metamaterials},
  author={Nguyen, Phu Thien and Heider, Yousef and Kochmann, Dennis M and Aldakheel, Fadi},
  journal={Computer Methods in Applied Mechanics and Engineering},
  volume={449},
  pages={118499},
  year={2026},
  doi={10.1016/j.cma.2025.118499},
  publisher={Elsevier}
}

@article{hsu2021microstructure,
  title={Microstructure Generation via Generative Adversarial Network for Heterogeneous, Topologically Complex {3D} Materials},
  author={Hsu, Tim and Epting, William K and Kim, Hokon and Abernathy, Harry W and Hackett, Gregory A and Rollett, Anthony D and Salvador, Paul A and Holm, Elizabeth A},
  journal={JOM},
  volume={73},
  number={1},
  pages={90--102},
  year={2021},
  doi={10.1007/s11837-020-04484-y},
  publisher={Springer}
}

@article{nguyen2022synthesizing,
  title={Synthesizing controlled microstructures of porous media using generative adversarial networks and reinforcement learning},
  author={Nguyen, Phong CH and Vlassis, Nikolaos N and Bahmani, Bahador and Sun, WaiChing and Udaykumar, HS and Baek, Stephen S},
  journal={Scientific reports},
  volume={12},
  number={1},
  pages={9034},
  year={2022},
  doi={10.1038/s41598-022-12845-7},
  publisher={Nature Publishing Group UK London}
}

@article{lyu2024microstructure,
  title={Microstructure reconstruction of {2D}/{3D} random materials via diffusion-based deep generative models},
  author={Lyu, Xianrui and Ren, Xiaodan},
  journal={Scientific Reports},
  volume={14},
  number={1},
  pages={5041},
  year={2024},
  doi={10.1038/s41598-024-54861-9},
  publisher={Nature Publishing Group UK London}
}

@article{baishnab20253d,
  title={{3D} multiphase heterogeneous microstructure generation using conditional latent diffusion models},
  author={Baishnab, Nirmal and Herron, Ethan and Balu, Aditya and Sarkar, Soumik and Krishnamurthy, Adarsh and Ganapathysubramanian, Baskar},
  journal={Digital Discovery},
  volume={4},
  number={11},
  pages={3175--3190},
  year={2025},
  doi={10.1039/d5dd00159e},
  publisher={Royal Society of Chemistry}
}

@article{ha2023rapid,
  title={Rapid inverse design of metamaterials based on prescribed mechanical behavior through machine learning},
  author={Ha, Chan Soo and Yao, Desheng and Xu, Zhenpeng and Liu, Chenang and Liu, Han and Elkins, Daniel and Kile, Matthew and Deshpande, Vikram and Kong, Zhenyu and Bauchy, Mathieu and others},
  journal={Nature Communications},
  volume={14},
  number={1},
  pages={5765},
  year={2023},
  doi={10.1038/s41467-023-40854-1},
  publisher={Nature Publishing Group UK London}
}

@article{zeng2023inverse,
  title={Inverse design of energy-absorbing metamaterials by topology optimization},
  author={Zeng, Qingliang and Duan, Shengyu and Zhao, Zeang and Wang, Panding and Lei, Hongshuai},
  journal={Advanced Science},
  volume={10},
  number={4},
  pages={2204977},
  year={2023},
  doi={10.1002/advs.202204977},
  publisher={Wiley Online Library}
}

@article{thillaithevan2024inverse,
  title={Inverse design of periodic microstructures with targeted nonlinear mechanical behaviour},
  author={Thillaithevan, Dilaksan and Murphy, Ryan and Hewson, Robert and Santer, Matthew},
  journal={Structural and Multidisciplinary Optimization},
  volume={67},
  number={4},
  pages={55},
  year={2024},
  doi={10.1007/s00158-024-03761-7},
  publisher={Springer}
}

@article{zhang2019computational,
  title={Computational design of finite strain auxetic metamaterials via topology optimization and nonlinear homogenization},
  author={Zhang, Guodong and Khandelwal, Kapil},
  journal={Computer Methods in Applied Mechanics and Engineering},
  volume={356},
  pages={490--527},
  year={2019},
  doi={10.1016/j.cma.2019.07.027},
  publisher={Elsevier}
}

@article{aveline2026topology,
  title={A topology optimization framework for the inverse design of nonlinear mechanical metamaterials},
  author={Aveline, Charlie and Santer, Matthew and Hewson, Robert},
  journal={Advanced Engineering Materials},
  volume={28},
  number={9},
  pages={e202503183},
  year={2026},
  doi={10.1002/adem.202503183},
  publisher={Wiley}
}

@article{alvarez2012kernels,
  title={Kernels for vector-valued functions: A review},
  author={Alvarez, Mauricio A and Rosasco, Lorenzo and Lawrence, Neil D},
  journal={Foundations and Trends{\textregistered} in Machine Learning},
  volume={4},
  number={3},
  pages={195--266},
  year={2012},
  doi={10.1561/2200000036},
  publisher={Emerald Publishing Limited}
}

@book{journel1976mining,
  title={Mining Geostatistics},
  author={Journel, A. G. and Huijbregts, C. J.},
  year={1978},
  publisher={Academic Press},
  address={London and New York},
  isbn={9780123910509}
}

@inproceedings{williams1995gaussian,
  title={Gaussian Processes for Regression},
  author={Williams, Christopher and Rasmussen, Carl},
  booktitle={Advances in Neural Information Processing Systems},
  editor={Touretzky, D. S. and Mozer, M. C. and Hasselmo, M. E.},
  volume={8},
  year={1995},
  publisher={MIT Press}
}

@inproceedings{titsias2009variational,
  title={Variational Learning of Inducing Variables in Sparse Gaussian Processes},
  author={Titsias, Michalis K.},
  booktitle={Proceedings of the Twelfth International Conference on Artificial Intelligence and Statistics},
  series={Proceedings of Machine Learning Research},
  volume={5},
  pages={567--574},
  year={2009},
  publisher={PMLR}
}

@inproceedings{hensman2013gaussian,
  title={Gaussian Processes for Big Data},
  author={Hensman, James and Fusi, Nicol{\`o} and Lawrence, Neil D.},
  booktitle={Proceedings of the Twenty-Ninth Conference on Uncertainty in Artificial Intelligence},
  pages={282--290},
  year={2013},
  publisher={AUAI Press}
}

@article{jordan1999introduction,
  title={An introduction to variational methods for graphical models},
  author={Jordan, Michael I and Ghahramani, Zoubin and Jaakkola, Tommi S and Saul, Lawrence K},
  journal={Machine learning},
  volume={37},
  number={2},
  pages={183--233},
  year={1999},
  doi={10.1023/a:1007665907178},
  publisher={Springer}
}

@book{bishop2006pattern,
  title={Pattern Recognition and Machine Learning},
  author={Bishop, Christopher M.},
  year={2006},
  publisher={Springer}
}

@article{kumar2020inverse,
  title={Inverse-designed spinodoid metamaterials},
  author={Kumar, Siddhant and Tan, Stephanie and Zheng, Li and Kochmann, Dennis M},
  journal={npj Computational Materials},
  volume={6},
  pages={73},
  year={2020},
  doi={10.1038/s41524-020-0341-6},
  publisher={Nature Publishing Group UK London}
}

@article{bastek2022truss,
  title={Inverting the structure--property map of truss metamaterials by deep learning},
  author={Bastek, Jan-Hendrik and Kumar, Siddhant and Telgen, Bastian and Glaesener, Rapha{\"e}l N. and Kochmann, Dennis M.},
  journal={Proceedings of the National Academy of Sciences},
  volume={119},
  number={1},
  pages={e2111505119},
  year={2022},
  doi={10.1073/pnas.2111505119},
  publisher={National Academy of Sciences}
}

@article{honarmandi2022batchbo,
  title={Accelerated materials design using batch Bayesian optimization: A case study for solving the inverse problem from materials microstructure to process specification},
  author={Honarmandi, P. and Attari, V. and Arroyave, R.},
  journal={Computational Materials Science},
  volume={210},
  pages={111417},
  year={2022},
  doi={10.1016/j.commatsci.2022.111417},
  publisher={Elsevier}
}

@article{dold2023differentiable,
  title={Differentiable graph-structured models for inverse design of lattice materials},
  author={Dold, Dominik and Aranguren van Egmond, Derek},
  journal={Cell Reports Physical Science},
  volume={4},
  pages={101586},
  year={2023},
  doi={10.1016/j.xcrp.2023.101586},
  publisher={Elsevier}
}

@article{dong2023blockcopolymerbo,
  title={Inverse design of complex block copolymers for exotic self-assembled structures based on Bayesian optimization},
  author={Dong, Qingshu and Gong, Xiangrui and Yuan, Kangrui and Jiang, Ying and Zhang, Liangshun and Li, Weihua},
  journal={ACS Macro Letters},
  volume={12},
  pages={401--407},
  year={2023},
  doi={10.1021/acsmacrolett.3c00020},
  publisher={American Chemical Society}
}

@article{zheng2023truss,
  title={Unifying the design space and optimizing linear and nonlinear truss metamaterials by generative modeling},
  author={Zheng, Li and Karapiperis, Konstantinos and Kumar, Siddhant and Kochmann, Dennis M.},
  journal={Nature Communications},
  volume={14},
  number={1},
  pages={7563},
  year={2023},
  doi={10.1038/s41467-023-42068-x},
  publisher={Nature Publishing Group}
}

@article{attari2023pfvae,
  title={Towards inverse microstructure-centered materials design using generative phase-field modeling and deep variational autoencoders},
  author={Attari, Vahid and Khatamsaz, Danial and Allaire, Douglas and Arroyave, Raymundo},
  journal={Acta Materialia},
  volume={259},
  pages={119204},
  year={2023},
  doi={10.1016/j.actamat.2023.119204},
  publisher={Elsevier}
}

@article{lininger2021thinfilmcnn,
  title={General inverse design of layered thin-film materials with convolutional neural networks},
  author={Lininger, Andrew and Hinczewski, Michael and Strangi, Giuseppe},
  journal={ACS Photonics},
  volume={8},
  number={12},
  pages={3641--3650},
  year={2021},
  doi={10.1021/acsphotonics.1c01498},
  publisher={American Chemical Society}
}

@article{mohammadnejad2024ann,
  title={Artificial neural networks for inverse design of a semi-auxetic metamaterial},
  author={Mohammadnejad, Mohammadreza and Montazeri, Amin and Bahmanpour, Ehsan and Mahnama, Maryam},
  journal={Thin-Walled Structures},
  volume={200},
  pages={111927},
  year={2024},
  doi={10.1016/j.tws.2024.111927},
  publisher={Elsevier}
}

@article{jadoon2025inverse,
  title={Inverse design of anisotropic microstructures using physics-augmented neural networks},
  author={Jadoon, Asghar A. and Kalina, Karl A. and Rausch, Manuel K. and Jones, Reese and Fuhg, Jan Niklas},
  journal={Journal of the Mechanics and Physics of Solids},
  volume={203},
  pages={106161},
  year={2025},
  doi={10.1016/j.jmps.2025.106161},
  publisher={Elsevier}
}

@article{maurizi2025graphmetamat,
  title={Designing metamaterials with programmable nonlinear responses and geometric constraints in graph space},
  author={Maurizi, Marco and Xu, Derek and Wang, Yu-Tong and Yao, Desheng and Hahn, David and Oudich, Mourad and Satpati, Anish and Bauchy, Mathieu and Wang, Wei and Sun, Yizhou and Jing, Yun and Zheng, Xiaoyu Rayne},
  journal={Nature Machine Intelligence},
  volume={7},
  number={7},
  pages={1023--1036},
  year={2025},
  doi={10.1038/s42256-025-01067-x},
  publisher={Springer Nature}
}

@article{mirzaee2025inverse,
  title={Inverse design of microstructures using conditional continuous normalizing flows},
  author={Mirzaee, Hossein and Kamrava, Serveh},
  journal={Acta Materialia},
  volume={285},
  pages={120704},
  year={2025},
  doi={10.1016/j.actamat.2024.120704},
  publisher={Elsevier}
}

@article{kang2025auxetic,
  title={Design of auxetic metamaterial for enhanced low cycle fatigue life and negative Poisson's ratio through multi-objective Bayesian optimization},
  author={Kang, Sukheon and Moon, Hyeonbin and Shin, Seonho and Mousavi, Mahmoud and Sung, Hyokyung and Ryu, Seunghwa},
  journal={Materials \& Design},
  volume={252},
  pages={113798},
  year={2025},
  doi={10.1016/j.matdes.2025.113798},
  publisher={Elsevier}
}

@article{zhang2025bio,
  title={Inverse design of nonlinear mechanics of bio-inspired materials through interface engineering and Bayesian optimization},
  author={Zhang, Wei and Tang, Mingjian and Mu, Haoxuan and Yang, Xingzi and Zeng, Xiaowei and Tuo, Rui and Chen, Wei and Gao, Wei},
  journal={Extreme Mechanics Letters},
  volume={78},
  pages={102359},
  year={2025},
  doi={10.1016/j.eml.2025.102359},
  publisher={Elsevier}
}

@article{zhang2025generative,
  title={Generative inverse design of metamaterials with customized stress-strain response},
  author={Zhang, Xin-Chun and Song, Zhi-Yi and Li, Yi-Nan and Xiao, Li-Jun and Xu, Zheng and Rao, Li-Xiang and Ci, Tie-Jun and Hui, Xu-Long},
  journal={International Journal of Mechanical Sciences},
  volume={306},
  pages={110875},
  year={2025},
  doi={10.1016/j.ijmecsci.2025.110875},
  publisher={Elsevier}
}

@article{zhang2025latticeoptdiff,
  title={Conditional diffusion models for the inverse design of lattice structures},
  author={Zhang, Jinlong and Chen, Shikun and Martin, Robert J. and Liu, Baochang and Zhang, Ruixiong and Xiao, Dengbao},
  journal={Structural and Multidisciplinary Optimization},
  volume={68},
  number={3},
  pages={58},
  year={2025},
  doi={10.1007/s00158-025-03984-2},
  publisher={Springer}
}

@article{garg2026gcnhelicoidal,
  title={Physics-guided graph convolutional network framework for inverse design of bioinspired helicoidal laminates},
  author={Garg, Aman and Shukla, Neeraj Kumar and Belarbi, Mohamed-Ouejdi and Nguyen, Tan N. and Raman, Roshan and Garg, Akhil},
  journal={Advanced Engineering Materials},
  pages={e202502703},
  year={2026},
  doi={10.1002/adem.202502703},
  publisher={Wiley}
}

@article{fan2026metaplate,
  title={An efficient inverse design framework of manufacturable phononic metaplates via combining generative deep learning and Bayesian latent space exploration},
  author={Fan, Lei and Su, Zhongqing},
  journal={Mechanical Systems and Signal Processing},
  volume={243},
  pages={113677},
  year={2026},
  doi={10.1016/j.ymssp.2025.113677},
  publisher={Elsevier}
}

@article{zang2026designgenno,
  title={Design-GenNO: A physics-informed generative model with neural operators for inverse microstructure design},
  author={Zang, Yaohua and Koutsourelakis, Phaedon-Stelios},
  journal={Computer Methods in Applied Mechanics and Engineering},
  volume={450},
  pages={118597},
  year={2026},
  doi={10.1016/j.cma.2025.118597},
  publisher={Elsevier}
}

@article{wang2020deep,
  title={Deep generative modeling for mechanistic-based learning and design of metamaterial systems},
  author={Wang, Liwei and Chan, Yu-Chin and Ahmed, Faez and Liu, Zhao and Zhu, Ping and Chen, Wei},
  journal={Computer Methods in Applied Mechanics and Engineering},
  volume={372},
  pages={113377},
  year={2020},
  doi={10.1016/j.cma.2020.113377},
  publisher={Elsevier}
}

@article{ma2019probabilistic,
  title={Probabilistic Representation and Inverse Design of Metamaterials Based on a Deep Generative Model with Semi-Supervised Learning Strategy},
  author={Ma, Wei and Cheng, Feng and Xu, Yihao and Wen, Qinlong and Liu, Yongmin},
  journal={Advanced Materials},
  volume={31},
  number={35},
  pages={1901111},
  year={2019},
  doi={10.1002/adma.201901111},
  publisher={Wiley Online Library}
}

@article{montes2018reduced,
  title={Reduced-order microstructure-sensitive models for damage initiation in two-phase composites},
  author={Montes de Oca Zapiain, David and Popova, Evdokia and Abdeljawad, Fadi and Foulk, James W and Kalidindi, Surya R and Lim, Hojun},
  journal={Integrating Materials and Manufacturing Innovation},
  volume={7},
  pages={97--115},
  year={2018},
  doi={10.1007/s40192-018-0112-0},
  publisher={Springer}
}

@article{olive2022characterization,
  title={Characterization of the symmetry class of an elasticity tensor using polynomial covariants},
  author={Olive, Marc and Kolev, Boris and Desmorat, Rodrigue and Desmorat, Boris},
  journal={Mathematics and Mechanics of Solids},
  volume={27},
  number={1},
  pages={144--190},
  year={2022},
  doi={10.1177/10812865211010885},
  publisher={SAGE Publications}
}

@article{riemer2026construction,
  title={Construction of minimal integrity bases for anisotropic hyperelasticity via structural tensors},
  author={Riemer, Brain M. and Brummund, J{\"o}rg and Kalina, Karl A. and Milor, Abel H. G. and Damma{\ss}, Franz and K{\"a}stner, Markus},
  journal={Journal of the Mechanics and Physics of Solids},
  volume={216},
  pages={106763},
  year={2026},
  doi={10.1016/j.jmps.2026.106763},
  publisher={Elsevier}
}

@software{danesh2026BayGDS,
  author    = {Danesh, Hooman and Wessels, Henning},
  title     = {Dataset for the publication "{Data}-efficient {Bayesian}-guided design selection from large candidate sets: Application to hyperelastic stochastic metamaterials"},
  year      = {2026},
  doi       = {10.5281/zenodo.19009893},
  publisher = {Zenodo}
}

\end{document}